\documentstyle[12pt,english,epsfig,cite]{article}
\title{The minimum width condition for neutrino conversion in matter}

\author{C.Lunardini$^{a)}$, A.Yu.Smirnov$^{b)}$}

\begin{document}

\def \lta {\mathrel{\vcenter{\hbox{$<$}\nointerlineskip\hbox{$\sim$}}}}
\def \gta {\mathrel{\vcenter{\hbox{$>$}\nointerlineskip\hbox{$\sim$}}}}

\bibliographystyle{unsrt}

\maketitle

\noindent
\begin{center}
{\small {\it a) SISSA-ISAS, via Beirut 2-4, 34100 Trieste, Italy} \\
{\it and INFN, sezione di Trieste, via Valerio 2, 34127 Trieste, Italy}\\
\vspace{0.4cm}
{\it b) The Abdus Salam ICTP, Strada Costiera 11, 34100 Trieste,
Italy} \\
{\it and Institute for Nuclear Research, RAS, Moscow, Russia}}
\end{center}
\vspace{0.5cm}

\begin{abstract}
\noindent
We find that for small vacuum mixing angle $\theta$ and low energies ($s\ll M^2_Z$)
the width of matter, $d_{1/2}$, needed to have conversion probability
$P\geq 1/2$ should be larger than  $d_{min}= \pi/(2\sqrt{2} G_{F}
\tan 2 \theta)$: $d_{1/2}\geq d_{min}$. 
Here $G_F$ is the Fermi constant, $s$ is the total energy squared in the center of
mass and $M_Z$ is the mass of the $Z$ boson. 
The absolute minimum $d_{1/2}=d_{min}$ is realized for oscillations in a uniform medium  
with resonance density. For all the other density distributions
(monotonically varying density, 
castle wall profile, etc.) the required width $d_{1/2}$ is larger
than $d_{min}$. 
The width $d_{min}$ depends on $s$, and for $Z$-resonance channels at $s\sim M^2_Z$ 
we get that $d_{min}(s)$ is 20 times smaller than the low energy value. 
We apply the minimum width condition, $d\geq d_{min}$, to
high energy neutrinos in matter as well as in neutrino background.
Using this condition, we conclude 
that the matter effect is negligible for neutrinos propagating in 
AGN and GRBs environments. Significant conversion can be expected for 
neutrinos crossing dark matter halos of clusters of galaxies
and for neutrinos produced by cosmologically distant
sources and propagating in the universe. 
\end{abstract}

\vskip 2cm
SISSA/13/2000/EP


\newpage

\section{Introduction}
Since the paper by Wolfenstein\cite{Wolfenstein:1978ue}, the   neutrino
transformations in matter became one of the most important phenomena in 
neutrino physics.  
Neutrinos propagating in matter undergo coherent forward scattering
(refraction) described at low energies by the potential 
\begin{equation}
V=\sqrt{2} G_{F}n~,
\label{eq:pot} 
\end{equation}
where $G_{F}$ is the Fermi
constant, and $n$ is a function of the density and chemical
composition of the medium.  For the case of $\nu_{e}-\nu_{\mu}$ and 
$\nu_{e}-\nu_{\tau}$  conversion in matter
$n$ coincides with the electron number density, $n_e$.
 
Refraction can lead to an enhancement of oscillations in media 
with constant density, and to resonant conversion in the varying density case 
(MSW effect)\cite{Mikheev:1986wj, MikSm}. 
For periodic, or quasi-periodic density profiles, various parametric effects
can occur\cite{Ermilova:1988pw, Akhmedov:1988kd,Krastev:1989ix}. 

The MSW effect has been applied to solar neutrinos\cite{Mikheev:1986wj}, 
and to neutrinos from supernovae\cite{Suzuki}.  
Oscillations of neutrinos of various origins (solar, atmospheric, 
supernovae neutrinos, etc.) in the matter of the Earth have been extensively 
studied. 
Apart from resonance enhancement of oscillations, parametric effects are
expected for neutrinos crossing both the mantle and  the core of the
Earth\cite{Liu:1998yb,Liu:1998nb,Petcov:1998su,Akhmedov:1999ui}. 
The oscillations and conversion of active neutrinos into a sterile species
can be important in the Early Universe\cite{Dolgov:1999wv}. 
Recently, matter effects on high energy neutrino fluxes from Active Galactic 
Nuclei (AGN) and  Gamma Ray Bursters (GRBs) have been estimated\cite{Waxman:1997hj}.  
Propagation of ultra-high energy neutrinos in halos of galaxies has been
considered\cite{Horvat:1998ym}. 

It is intuitively clear that to have a significant matter effect a sufficiently
large amount of matter is needed.
Let us define the width of the medium as the integrated  
density along the path travelled by the neutrino in the matter:
\begin{equation}
d = \int n_e (L) dL~.
\label{eq:ddef} 
\end{equation}
This quantity is frequently named  ``column density" in astrophysical context.  
We will show that there exists a minimum
value  $d_{min}$ for the width below which it is not possible to have 
significant neutrino conversion. This lower bound is independent of the 
density profile and of the neutrino energy and mass. 
That allows us to make conclusions on the relevance of matter effect in 
various situations without knowledge of the density distribution.

The paper is organized as follows: in section 2 we derive the minimum width
condition for the conversion in matter between two active neutrino
flavours, and check it for different density profiles. In section 3 we
discuss the generalizations of the condition to the active-sterile case
and to conversion induced by flavour changing neutrino-matter interactions.
We also study the matter effect in the small width limits. 
Section 4 presents a study of the minimum width condition for high energy 
neutrinos both in  matter and in neutrino background.
Section 5 is devoted to applications of our results to neutrino propagation
in AGN and GRBs environments, in dark matter halos and in the Early Universe. 
Conclusions and discussion follow in section 6. 

 
\section{The minimum width condition}
In this section we consider various mechanisms of matter enhancement of 
neutrino flavour conversion. 
For each of them we work out the minimum width of the medium needed to have
significant conversion probability, showing that a lower bound for the width
exists and is realized in the case of uniform medium with resonance density.


\subsection{The absolute minimum width}
Let us consider a system of two mixed flavour states\footnote{
The arguments remain the same for three neutrinos.}, $\nu_{e}$ and $\nu_{\mu}$ 
($\nu_{\tau}$), characterized by vacuum mixing angle $\theta$ and mass
squared difference $\Delta m^2$.
In a uniform medium the states oscillate and the transition  probability, 
as a  function of the distance $L$, is given by:
\begin{equation}  
P_{\nu_{e}\rightarrow \nu_{\mu}}(L)=\sin^2 2 \theta_m 
\sin^2\left(\pi {{L}\over{l_m}}\right),      
\label{eq:pconv} 
\end{equation}
where $\theta_m$ and $l_m$ are the mixing angle and the oscillation
length in the medium:
\begin{eqnarray}  
\sin 2 \theta_m= {\sin 2 \theta\over [(2E V/\Delta m^2 -\cos 2 \theta)^2+\sin^2 2 \theta]^{1/2}} \nonumber \\
l_m = {l\over [(2E V/\Delta m^2 -\cos 2 \theta)^2+\sin^2 2 \theta]^{1/2}}~.
\label{eq:form}
\end{eqnarray}
Here $l=4\pi E/\Delta m^2$ is the vacuum oscillation length and 
$E$ is the neutrino energy.

We assume that the vacuum mixing is small, so that vacuum oscillations effects 
are negligible ($P^{vac}_{\nu_{e}\rightarrow \nu_{\mu}} \ll 1$) and a strong 
transition in medium, i.e. $P_{\nu_{e}\rightarrow \nu_{\mu}}= O(1)$, 
is essentially due to matter effect. 
For definitness, we choose the condition of significant conversion to be
\begin{equation}
P_{\nu_{e}\rightarrow \nu_{\mu}} \geq {1 \over 2}~.
\label{eq:half} 
\end{equation}

Let us consider a uniform medium with resonance density\cite{Mikheev:1986wj}:
\begin{equation}  
n^{res}_e = {\Delta m^2 \over 2\sqrt{2}E G_{F}} \cos 2 \theta~.      
\label{eq:nres} 
\end{equation}
In this case the oscillation amplitude is $\sin 2\theta_m=1$, 
and the oscillation length equals  
\begin{eqnarray}  
l^{res} = {l\over \sin 2 \theta}={4\pi E \over \Delta m^2 \sin 2 \theta}~.
\label{eq:lres}
\end{eqnarray}
According to eq. (\ref{eq:pconv}) the condition (\ref{eq:half}) starts
to be satisfied for $L={l_m/ 4}$, and the corresponding width is:
\begin{equation}
d_{min}= {1 \over 4} n^{res}_e l^{res}.
\label{eq:d01} 
\end{equation}
Inserting the expressions of $n^{res}_e$ and $l^{res}$ given in
(\ref{eq:nres}) and (\ref{eq:lres}), we get:
\begin{eqnarray}
d_{min}= {\pi \over 2\sqrt{2} G_{F}\tan 2\theta}={d_0\over\tan 2 \theta}~,  
\label{eq:dmin} 
\end{eqnarray}
where 
\begin{eqnarray}
d_0={\pi \over 2\sqrt{2} G_{F}}\simeq{1.11\over G_{F}}~. 
\label{eq:d0} 
\end{eqnarray}
We will call $d_0$ the refraction width.
Numerically,
\begin{equation}
d_0=2.45 \cdot 10^{32} {\rm cm^{-2}}=4.08 \cdot 10^{8}\thinspace 
{\rm A\thinspace cm^{-2}}~,
\label{eq:d0num} 
\end{equation}
where $A=6\cdot 10^{23}$ is the Avogadro number\footnote{It can be 
checked that different choices of the condition (\ref{eq:half}) lead to 
analogous results. For instance, taking 
$P_{\nu_{e}\rightarrow \nu_{\mu}} \geq {3 \over 4}$ we find 
$$
d^{3/4}_0={4\over 3}d_0={2 \pi \over 3\sqrt{2} G_{F}}=5.41 \cdot 10^{8}
\thinspace {\rm A\thinspace cm^{-2}},
$$
and, for
$P_{\nu_{e}\rightarrow \nu_{\mu}} \geq {1 \over 4}$: 
$$
d^{1/4}_0= {2\over 3}d_0={\pi \over 3\sqrt{2} G_{F}}=2.7 \cdot 10^{8}
\thinspace {\rm A\thinspace cm^{-2}}.
$$
}. 

The widths $d_{min}$ and $d_0$ have a simple physical interpretation.
The refraction width $d_0$ is a universal quantity: 
it is  determined only by the Fermi coupling constant, and does not depend 
on the neutrino parameters at all.
Using the definition of refraction length 
\begin{equation}
l_0\equiv{2\pi\over V}={2\pi\over\sqrt{2}n_e G_{F}}
\label{eq:l0} 
\end{equation}
we can write:
\begin{equation}
{d_0\over n_e}={l_0\over 4}~.
\label{eq:dref} 
\end{equation}
It appears that $d_0$ corresponds to the distance at which the matter-induced 
phase difference between the flavour states equals $\pi/2$.
This can be considered as the definition of refraction width, 
which by eq. (\ref{eq:l0}) can be written in the general form:
\begin{equation}
d_0\equiv{\pi \over 2}{n_m \over V_m}~,
\label{eq:d0def} 
\end{equation}
where $V_m$ is the neutrino-medium potential and $n_m$ is the concentration of 
the relevant scatterers in the medium.

The minimum width, $d_{min}$, is inversely proportional to $\tan 2\theta$, 
which represents properties (the mixing) of the neutrino system itself.  
The smaller the mixing $\theta$, the larger is the width $d_{min}$  
needed for strong transition.

The condition (\ref{eq:half}) can be generalized. It corresponds to the case 
of initial state coinciding with a pure flavour state.  
In general one can require that the change
of the probability to detect a given flavour $\alpha$ is larger than $1/2$:
\begin{equation}
\Delta P\equiv P_f(\nu_\alpha)-P_i(\nu_\alpha)\geq {1\over 2}~, 
\label{eq:deltapgen} 
\end{equation}
where $P_i$ and $P_f$ are the initial and final probabilities.
The condition (\ref{eq:half}) corresponds to $P_i(\nu_\mu)=0$, so that 
$P_f(\nu_\alpha)=P_{\nu_{e}\rightarrow \nu_{\mu}}$.  
Taking $P_i=1/4$ and $P_f=3/4$,  we get in  a similar way:
\begin{eqnarray}
d_{1/2}={2\over 3}d_{min}= {\pi \over 3\sqrt{2} G_{F}\tan 2\theta}~. 
\label{eq:dmingen} 
\end{eqnarray}
This $d_{1/2}$ is the extreme value, however for most practical situations 
the condition (\ref{eq:half}) is more relevant, 
and from here on we will use the the width $d_{min}$ determined in 
(\ref{eq:dmin}).

In what follows we will show that for all the other density profiles the
width $d_{1/2}$ required by the condition (\ref{eq:half}) is larger than 
$d_{min}$. 


\subsection{Uniform medium with density out of resonance}
For $n_e \neq n^{res}_e$ the inequality (\ref{eq:half}) can be satisfied 
only if $\sin^2 2 \theta_m \geq {1 \over 2}$, which means that the density 
is required to be in the resonance interval: 
$n^{res}_e (1-\tan 2\theta) \leq n_e \leq n^{res}_e (1+\tan 2\theta)$.
At the edges of the interval we get the width
\begin{equation} 
d_{1/2}={\pi \over 2 G_{F}} \left({1\over \tan 2 \theta} \pm 1\right) \simeq 
\sqrt{2} d_{min}~, 
\label{eq:dbord} 
\end{equation}
which is larger than $d_{min}$.  
For other values of the density in the resonance interval we have $d_{min}<d_{1/2}<\sqrt{2} 
d_{min}$\footnote{ 
It can be checked that the width $d_{1/2}$ in 
eq. (\ref{eq:dbord}) is larger than $d_{min}$ for small mixing: 
$\sin 2\theta\lta 0.3$. 
We will use this condition as criterion of smallness of the mixing.}. 

 
\subsection{ Medium with varying density}
In general the neutrino propagation has a character of interplay 
of resonance conversion and oscillations.
Two conditions are needed for strong transition: 
\\

\noindent  
1)\thinspace Resonance condition: the neutrinos should cross the layer with
resonance density.
\\

\noindent
2)\thinspace Adiabaticity condition: the density should vary slowly enough. 
This condition can be written in terms of the adiabaticity parameter $\gamma$ 
at resonance:
\begin{eqnarray}
&&\gamma \ll 1 
\label{eq:gamma} \\
&&\gamma\equiv{2 E \cos 2\theta \over \Delta m^2 \sin^2 2 \theta}{1\over
n_e}\left|{dn_e\over dL}\right|~.
\label{eq:adiab} 
\end{eqnarray}
Notice that both the conditions 1) and 2) are local, and can be fulfilled for 
arbitrarily small widths of the medium. 
Clearly, they are not sufficient to assure a significant
conversion, and a third condition of large enough matter width is needed.  

Let us consider a linear density profile with length $2L$ and average density
equal to the resonance one, so that $n_{max}=n^{res}_e + \Delta n$ and 
$n_{min}=n^{res}_e - \Delta n$.    
Denoting $\theta_{1 m}$ and $\theta_{2 m}$ the mixings in the initial and final 
points, we find that in the first order of adiabatic perturbation theory 
the conversion probability is given by:  
\begin{eqnarray}  
P_{\nu_{e}\rightarrow \nu_{\mu}}(L)&=&
{1\over 2}-{1\over 2}\cos 2\theta_{1 m}\cos 2\theta_{2 m}  \nonumber \\
&-& {1\over 2}\sin 2\theta_{1 m}\sin 2\theta_{2 m} 
\cos\left[{1\over \gamma}f(x)\right]\nonumber \\
&-&2\sin (2\theta_{1 m}-2\theta_{2 m}) \alpha(x)
\cos\left[{1\over 2\gamma}f(x)\right]~, 
\label{eq:pconvad}
\end{eqnarray}
where
\begin{eqnarray}
x&=&2\pi \gamma {L\over l^{res}}~, \nonumber \\
f(x)&=&\ln(x+\sqrt{1+x^2})+x \sqrt{1+x^2}~,  \nonumber \\
\alpha(x)&=&\int^{x}_0 {dy\over 1+y^2}\cos\left[{1\over 2\gamma}f(y)\right]~.  
\label{eq:funct} 
\end{eqnarray}
For $\gamma\ll 1$ we get from eqs. (\ref{eq:pconvad}) and (\ref{eq:funct}):
\begin{equation}
d_{1/2}=d_{min} \left[1+ \left(1-{\pi \over 8}\right)\gamma^2\right].
\label{eq:pert} 
\end{equation}
This expression shows that for the adiabatic case $d_{1/2}\simeq d_{min}$ and
for weak violation of adiabaticity the
minimum width increases quadratically with  $\gamma$.  
We remark that in this case the effect is dominated by
oscillations with large (close to maximal) depth.  The change of density gives
only small corrections. 

Let us consider now a situation in which the resonance adiabatic conversion
is the main mechanism of flavour transition. A pure conversion effect is
realized if the initial neutrino state that enters the medium coincides with one
of the eigenstates of the Hamiltonian in matter, and the propagation in matter
is adiabatic. 
In this case no phase effect, and therefore no
oscillations occur.  Let us denote $n_i$ and $n_f$ the initial and final
densities of the medium, and suppose the initial state is 
$\nu_i=\nu_{2 m}=\sin \theta_m \nu_e+\cos \theta_m \nu_\mu$. 
The probability to find a $\nu_\mu$ in this state is $P_i(\nu_\mu)=
\cos^2\theta_m(n_i)$. The state evolves following the change of density,
so that it remains an eigenstate of the Hamiltonian, and 
the probability to find $\nu_\mu$ in the final state is $P_f(\nu_\mu)=
\cos^2\theta_m(n_f)$.  
Since the initial state $\nu_i$ does not coincide with a pure flavour state 
we will use the condition (\ref{eq:deltapgen}) as criterion of strong matter 
effect.
Inserting $P_i$ and $P_f$ in (\ref{eq:deltapgen}), 
we get the condition for $d_{1/2}$:
\begin{equation}
\cos 2\theta_m(n_f)-\cos 2\theta_m(n_i)=1~.
\label{eq:deltacos} 
\end{equation} 
Taking the initial and final values of the density as 
$n_i=n^{res}_e + \Delta n$ and 
$n_f=n^{res}_e - \Delta n$ ($\Delta n \geq 0$), and using the definition 
(\ref{eq:form}) we find that the equality (\ref{eq:deltacos}) leads to
\begin{equation}
\Delta n= n^{res}_e {1\over \sqrt{3}}\tan 2\theta~.
\label{eq:deltan} 
\end{equation} 
Clearly, for a given $\Delta n$ the size of the layer, 
and therefore its width, depend on the gradient of the density which can be
expressed in terms of the adiabaticity parameter $\gamma$, eq. (\ref{eq:adiab}).
We get:
\begin{equation}
n_e (L) dL={2E \cos 2\theta \over \Delta m^2 \sin^2 2\theta}{1\over \gamma}
dn_e~, 
\label{eq:adiabintdif} 
\end{equation}
and then integrating this equation we obtain:
\begin{equation}
d ={2E \cos 2\theta \over \Delta m^2 \sin^2 2\theta}{1\over \gamma}\Delta n~. 
\label{eq:adiabint} 
\end{equation}
Finally, inserting the expressions (\ref{eq:deltan}) and (\ref{eq:nres})
 in eq. (\ref{eq:adiabint}) we find:
\begin{equation}
d_{1/2} ={4\over \pi \sqrt{3}}{1\over \gamma} d_{min}~.  
\label{eq:mind12} 
\end{equation}
Let us comment on this result.  As far as the adiabaticity condition is
satisfied, the change of probability does not depend on the density
distribution; it is a function of the initial and final
densities only. 
If $\Delta n$ is fixed, the decrease of the width means the decrease of the
length $L$ of the layer, and therefore increase of the gradient of the density.
This will lead eventually to violation of the adiabaticity condition.  Thus,
the minimal width corresponds to the maximal $\gamma$ for which the
adiabaticity is not broken substantially.

For strong adiabaticity violation an increase of $d_{1/2}$ is expected, 
due to the increase of the minimum $\Delta n$ required by the condition 
(\ref{eq:deltapgen}), and therefore of the corresponding length. 
This can be seen if we consider the previous argument taking into account the
effect of the adiabaticity breaking from the beginning. 
Using the Landau-Zener level crossing probability 
$P_{LZ}=\exp(-{\pi/ 2\gamma})$,
which describes the transition between two eigenstates, we get, instead of 
(\ref{eq:deltacos}):
\begin{equation}
(1-2P_{LZ})(\cos 2\theta_m(n_f)-\cos 2\theta_m(n_i))=1~,
\label{eq:deltacoslz} 
\end{equation}
where we have averaged out the interference terms.    
Then instead of eq. (\ref{eq:deltan}) we get
\begin{equation}
\Delta n = n^{res}_e {1\over \sqrt{16 P^2_{LZ}-16P_{LZ}+3}}\tan 2\theta~.
\label{eq:deltanlz} 
\end{equation} 
Finally, the condition for $d_{1/2}$ can be written as:
\begin{equation}
d_{1/2} ={4\over \pi \gamma \sqrt{16 P^2_{LZ}-16P_{LZ}+3}} d_{min}~.  
\label{eq:mind12lz} 
\end{equation}
For $\gamma\rightarrow 0$ eq. (\ref{eq:mind12lz}) gives 
$d_{1/2}\rightarrow \infty$, according to the fact that the density changes 
very slowly and therefore the width needed to have significant conversion 
increases.  
With the increase of $\gamma$ the width $d_{1/2}$ decreases and  
has a minimum at $\gamma \simeq$0.7,
for which we find $d_{1/2}\simeq 1.5 \thinspace d_{min}$. 
With further increase of $\gamma$  ($\gamma \gta 0.7$) the width $d_{1/2}$
increases rapidly. According to (\ref{eq:mind12lz}) it diverges for 
$P\rightarrow 1/4$, when $\gamma \rightarrow \pi /(4 \ln 2)\simeq 1.13$.   
This value corresponds to the case in which
the adiabaticity violation is so strong that even an infinite amount of matter
is not enough to satisfy the condition (\ref{eq:deltapgen}).
Thus, we have found that also in this case $d_{1/2}>d_{min}$.


\subsection{Step-like profile}
As an extreme case of strong adiabaticity 
violation, let us consider the profile consisting of two layers of
matter, having densities $n_1=n^{res}_e + \Delta n$ and 
$n_2=n^{res}_e - \Delta n$ ($\Delta n \geq 0$), and equal lengths $L_1=L_2=L$.
At the border between the layers the density has a jump of size $2\Delta n$.
We fix $L={l^{res}/ 8}$, so that $d=d_{min}$. 
The result for the conversion probability can be computed exactly:
\begin{equation}
P^{step}_{\nu_{e}\rightarrow \nu_{\mu}}=s^2 
\sin^2\left({{\pi}\over{4 s}}\right)
+s^2 c^2 \left[1-\cos\left({{\pi}\over{4 s}}\right)\right]^2~,
\label{eq:step}
\end{equation}
where we denote the mixing parameters in the two layers as 
$\sin 2\theta_{2 m}=\sin 2\theta_{1 m} \equiv s$, 
$\cos 2\theta_{2 m}=-\cos 2\theta_{1 m}\equiv c$. 
In absence of the step ($\Delta n=0$), 
$P^{step}_{\nu_{e}\rightarrow \nu_{\mu}}$  equals $1/2$, 
recovering the case $n_e=n^{res}_e={\rm const}$. 
The probability (\ref{eq:step}) decreases monotonically
as $\Delta n$ increases. 
Expanding in $\delta=({\Delta n / n^{res}_e \tan 2 \theta})^2$ 
we get\footnote{This approximation proves to be very good (relative error 
$ \le 0.5 \%$) for $0 \le \delta \le 1$, i.e. for  $n_1$ and $n_2$ in the 
resonance interval.}: 
\begin{eqnarray}
P^{step}_{\nu_{e}\rightarrow \nu_{\mu}} \simeq
{1\over 2}-(\sqrt{2}-1-{\pi \over 8})\delta  \simeq 
{1\over 2}-0.02\thinspace \delta~. 
\label{eq:exp} 
\end{eqnarray}
According to (\ref{eq:exp}), for $d=d_0$ we have 
$P^{step}_{\nu_{e}\rightarrow \nu_{\mu}}<1/2$. This implies that, to have 
$P^{step}_{\nu_{e}\rightarrow \nu_{\mu}}=1/2$ one needs
$d_{1/2}>d_{min}$.


\subsection{ Castle-wall profile}
The profile consists of a  periodical sequence of alternate layers of
matter, having two different densities $n_1$ and $n_2$.
We denote the corresponding mixing angles as 
$\theta_{1 m}$ and $\theta_{2 m}$.
In this case, a strong transformation  requires certain 
conditions on the oscillation 
phases acquired by neutrinos in the layers\cite{Akhmedov:1999ui};  
therefore the transformation is a consequence of the specific density profile,
rather than of  an enhancement of the mixing.
Suppose $n_1 \ll n^{res}_e$ and $n_2=0$, and 
take the width of each layer to be equal to half oscillation
length, so that the oscillation phase acquired in each layer is $\pi$.  
It can be shown\cite{Krastev:1989ix} that for small $\theta$ this is the 
condition under which the conversion probability
increases most rapidly with the distance. 
As a function of the number $N$ of periods (a period corresponds to two
layers), the probability is given by\cite{Krastev:1989ix,Akhmedov:1988kd}:
\begin{equation}  
P_{\nu_{e}\rightarrow \nu_{\mu}}(N)= \sin^2(2 N \Delta \theta)~,   
\label{eq:pcastle} 
\end{equation}
where $\Delta \theta\equiv\theta_{1 m}-\theta_{2 m}=\theta_{1 m}-\theta$.
Using the approximation $2\theta_{1 m}-2\theta\simeq 
\sin 2\theta_{1 m}-\sin 2\theta$, and expanding $\sin 2\theta_{1 m}$ in $n_1$, 
we get: 
\begin{equation}
d_{1/2}={\pi^2 \over 2\sqrt{2} G_{F}}{1\over \sin 2 \theta} \simeq \pi d_{min}~.
\label{eq:castle} 
\end{equation}
Again, we find that $d_{1/2}\ge d_{min}$.
\\

Thus, for all the known mechanisms of matter enhancement of flavour transition 
(resonant oscillations, adiabatic conversion, parametric effects),
we have found that the width $d_{1/2}$ is larger than $d_{min}$, which is 
realized for the case of uniform medium with resonance density.
In fact the constant profile with resonance density could be expected from the 
beginning to represent an extreme case: 
this profile is singled out, since  it is the simplest distribution with the 
density fixed at the unique value $n^{res}_e$. 

It is worthwile to introduce also the total nucleon width. 
Let us consider a medium made of electrons, protons and
neutrons with number densities $n_e$, $n_p$  and $n_n$.
Defining the number of electrons per nucleon as $Y_e\equiv n_e/(n_n+n_p)$,
we can write the total nucleon width
that corresponds to $d_0$ as:
\begin{equation}
d_{0 N}\equiv {d_0\over Y_e}~.
\label{eq:d0N} 
\end{equation}
We can also introduce the total mass width $d_{\rho}$:
\begin{equation}
d_\rho\equiv m_N d_{0 N}={m_N d_0\over Y_e}~.
\label{eq:drho} 
\end{equation}
For electrically and isotopically neutral medium $(n_e=n_n=n_p)$, 
eq. (\ref{eq:d0N}) gives:
\begin{equation}
d_{0 N}=2 d_0~,
\label{eq:2d0N} 
\end{equation}
and numerically:
\begin{equation}
d_{0 N}= 4.9 \cdot 10^{32} {\rm cm^{-2}}~,  \hskip 0.9cm 
{d_{\rho} }=8.16 \cdot 10^8 ~{\rm g \cdot cm^{-2}}~.
\label{eq:d0numord} 
\end{equation}


\section{Generalizations}
In this section we generalize the previous results to active-sterile transition
and to the case of flavour-changing induced
conversion.  We also discuss the  small width limits. 


\subsection{Active-sterile conversion}
In this case the scattering both on electrons and on nucleons contributes 
to the conversion, and the effective potential for an electron neutrino in an 
electrically neutral medium equals 
\begin{equation}
V=\sqrt{2} G_{F}n_e\left( 1-{n_n\over 2 n_p}\right).
\label{eq:potst} 
\end{equation} 
Thus, the results for $\nu_e-\nu_s$ transition can be obtained from
those for $\nu_e-\nu_\mu$  by the replacement
$n_e \rightarrow n_e ( 1-{n_n / 2 n_p})$.  For the refraction width we get 
immediately:
\begin{equation}
d_0(\nu_e \rightarrow \nu_s) = d_0 \left|{ 2 n_p \over 
 2 n_p-n_n }\right|~. 
\label{eq:dstergen}
\end{equation}
In particular, for an isotopically neutral medium eq. (\ref{eq:dstergen}) 
gives
\begin{equation}
d_0(\nu_e \rightarrow \nu_s) = 2 d_0~.
\label{eq:dster}
\end{equation}
Notice that, for highly neutronized media ($n_n \gg n_p$) the width  
 $d_0(\nu_e \rightarrow \nu_s)$ gets significantly smaller than $d_0$. 
In this case, however, the physical situation is more properly described by 
the total nucleon width $d_{0 N}$ defined in eq. (\ref{eq:d0N}), since the 
effect is mainly due to the scattering on neutrons.
We find:
\begin{equation}
d_{0 N}(\nu_e \rightarrow \nu_s) =2 d_0
\left|{n_p+n_n \over 2 n_p-n_n}\right|~, 
\label{eq:drhoster}
\end{equation}
which gives in the limit $n_n\rightarrow \infty$:
\begin{equation}
d_{0 N}(\nu_e \rightarrow \nu_s) = 2 d_0~,
\label{eq:dsterhn}
\end{equation}
similarly to eq. (\ref{eq:dster}).

For the  $\nu_\mu-\nu_s$ case, the potential, and consequently the width, 
can be obtained by replacing $n_e \rightarrow n_e\left(-{n_n/ 2 n_p}\right)$, 
which gives 
\begin{equation}
d_0(\nu_\mu \rightarrow \nu_s) = d_0 {2 n_p\over n_n}~.
\label{eq:dstermu}
\end{equation}
For isotopically neutral medium, eq. (\ref{eq:dstermu}) reduces to
eq. (\ref{eq:dster}).  For highly neutronized media, the argument is analogous
to the one for the $\nu_e-\nu_s$ case, and we get the same result as in  
eq. (\ref{eq:dsterhn}).


\subsection{Oscillations induced by flavour changing (FC) neutrino-matter 
interactions}
In this case the neutrino masses can be zero, or negligible, and the flavour
transition is a pure matter effect. 
The Hamiltonian of the system has the following form\cite{Wolfenstein:1978ue}:
\begin{equation}
H=\sqrt{2} G_F \pmatrix{0 & \epsilon n_f \cr 
\epsilon n_f & \epsilon^\prime n_f \cr},
\label{eq:fch}
\end{equation}
where $n_f$ is the effective number density of the scatterers,
and  $\epsilon$ and $\epsilon^\prime$ are parameters of the interaction.

As follows from eq. (\ref{eq:fch}), in a uniform medium
the neutrinos oscillate with transition probability:
\begin{eqnarray}  
&&P_{\nu_{e}\rightarrow \nu_{\mu}}(L)={4\epsilon^2\over 4\epsilon^2+
\epsilon^{\prime 2} }
\sin^2\left(\pi {{L}\over{l}}\right), \\  
&&l={\pi \sqrt{2}\over \sqrt{4\epsilon^2+\epsilon^{\prime 2}}}{1\over G_F n_f}~.    
\label{eq:pconvfc} 
\end{eqnarray} 
We assume
$\epsilon^\prime < \epsilon$, which is needed to have  a significant 
oscillation amplitude.
Using eq. (\ref{eq:d01}), we get:
\begin{eqnarray}
d^{FC}_{1/2}&=& {\pi \over 2\sqrt{2} G_{F}} {1\over\sqrt{4\epsilon^2+
\epsilon^{\prime 2}}}
{n_e\over n_f}  \nonumber \\
 &=&{d_{0} \over\sqrt{4\epsilon^2+\epsilon^{\prime 2}}}
{n_e\over n_f}~.
\label{eq:d0fc} 
\end{eqnarray}

Notice that the factor 
$(n_e/ n_f)\tan 2\theta/ \sqrt{4\epsilon^2+\epsilon^{\prime 2}}$ 
implies that $d^{FC}_{1/2}$ can be significantly smaller than $d_{min}$, 
and the oscillation effect can be observed in media of smaller width.
For a FC neutrino interaction with up (or down) quarks and isotopically 
neutral medium we have ${n_e/n_f}=1/3$, and therefore:
\begin{equation}
d^{FC}_{1/2}\simeq {1\over 6\epsilon}d_{0}~.
\label{eq:d0fc6} 
\end{equation}
Notice that there are two sources of decrease of the width: 
the factor $2$ is given by the presence of
two off diagonal terms in the Hamiltonian (\ref{eq:fch}) and the factor $3$ is
due to the larger number of scatterers. 
Taking  $\epsilon=1$, we find the value: 
\begin{equation}
d^{FC}_{1/2}\simeq 1.36 \cdot 10^8 \thinspace{\rm g \cdot cm^{-2}}.
\label{eq:d0fcnum} 
\end{equation}
For a density $n=4 \thinspace {\rm g \cdot cm^{-3}}$ (Earth's crust), this 
corresponds to the distance $L= 337 \thinspace{\rm Km}$, 
which is comparable to the length of the present long base-line neutrino
experiments: 
K2K  (base-line 250 Km) and ICANOE (740 Km).


\subsection{The small width limits}
In a number of situations (see section 5)
the width of the medium is smaller, or much smaller, than $d_{min}$. 
We consider, then, the matter effect on oscillations in the limit 
$d/d_{min} \ll 1$.
Introducing the two variables:
\begin{eqnarray} 
\lambda\equiv {L \over l^{res}/4} \hskip 0.9cm
\rho\equiv{n_e \over n^{res}_e}~,
\label{eq:twodef}
\end{eqnarray}
we can write the small width condition as:
\begin{equation} 
d/d_{min}=\lambda \rho\ll 1.
\label{eq:smalld}
\end{equation} 

We focus on two important realizations of this inequality:
\\

\noindent  
1) Small size of the layer and density close to resonance. 
As we have shown in section 2, a strong transition requires 
$n_e\simeq n^{res}_e$. In case of small width this implies small size 
of the layer. 
Therefore we have $\lambda \ll 1$ and 
$\rho\sim 1$. In this case $d/d_{min}\simeq \lambda$.  
We expand the oscillation probability (\ref{eq:pconv}) in $\lambda$ 
at the lowest (nonzero) order, and  find that the matter effect vanishes 
quadratically with $d/d_{min}$:    
\begin{equation} 
P(\lambda,\rho)\simeq
\left({\pi \over 4}\right)^2 \lambda^2 \sim 
\left({\pi \over 4}\right)^2\left({d\over d_{min}}\right)^2~.  
\label{eq:smallsize}
\end{equation}   
\\

\noindent
2) Small density of the medium and length close to the minimum 
value 
$l^{res}/4$.
Another condition of strong conversion is to have the size of the layer of the 
order of the oscillation length. 
According to eq. (\ref{eq:smalld}), this means that the density is small.
Thus, we have $\rho \ll 1$ and $\lambda\sim 1$, and therefore 
$d/d_{min}\simeq \rho$.  
In order to give a phase-independent description 
of the matter effect, we perform an expansion in $\rho$ of the
oscillation amplitude:
\begin{eqnarray} 
\sin^2 2\theta_m-\sin^2 2\theta=2\rho \sin^2 2\theta \cos 2\theta 
\sim 2{d\over d_{min}}\sin^2 2\theta \cos 2\theta ~.
\label{eq:smalln}
\end{eqnarray} 
Unlike the previous case, the relative matter effect is linear in $d/d_{min}$.


\section{ Refraction of high energy neutrinos}
In this section we examine the refraction of high energy neutrinos 
($s\gta M^2_Z$), both in matter and in neutrino background.  

\subsection{High energy neutrinos in matter}
Let us consider the propagation of high energy neutrinos in medium composed of protons,
neutrons and electrons. 
The expressions (\ref{eq:pot}) and (\ref{eq:d0}) refer to the low energy range, 
$s \ll M^2_W$, where $s$ is the center of mass energy squared
of the incoming neutrino and the target electron, 
and $M_W$ is the mass of the $W$ boson. 
The general formulas, valid for high energies too, can be obtained  
by restoring the effect of the complete propagator of
the $W$ boson in the expression of the potential (an analogous argument holds for the
$Z$ boson). 

Let us consider $\nu_e- \nu_\mu$ conversion.
Since the refraction effects are determined by the real part of the propagator, the
potential (\ref{eq:pot}) is generalized as:
\begin{eqnarray}
V&=&\sqrt{2} G_{F}n_e f(q^2_W) 
\label{eq:replpot} \\
f(q^2_W)&\equiv&{1-q^2_W \over (1-q^2_W)^2+\gamma^2_W }~, 
\label{eq:ffunc}
\end{eqnarray}
where $q^2_W\equiv q^2/M^2_W$ and $\gamma_W\equiv \Gamma_W/M_W$;
$q$ and $\Gamma_W$ are the four momentum and the width of the $W$ boson.

The only contribution to the potential (\ref{eq:replpot}) is given by the forward charged current 
scattering on electrons ($u$-channel exchange of $W$), 
for which $q^2\simeq -s$. Therefore, introducing $s_W\equiv s/M^2_W$,
we have $q^2_W\simeq -s_W$.
From the potential (\ref{eq:replpot}) we can find the refraction width 
$d_0$ using the definition (\ref{eq:d0def}).  
For the nucleon refraction width (\ref{eq:d0N}) we find: 
\begin{equation}
d_{0 N}(s_W)= {1\over Y_e}{d_0\over f(-s_W)}={d_0\over Y_e}(1+s_W)~,
\label{eq:repl}
\end{equation}
where we have neglected the width $\gamma_W$.
Eq. (\ref{eq:repl}) shows that $d_{0 N}(s_W)$, and therefore $d_{min}(s_W)$, increase
linearly with $s_W$ above the threshold  of the $W$ boson
production. 

\begin{figure}[hbt]
\begin{center}
\epsfig{file=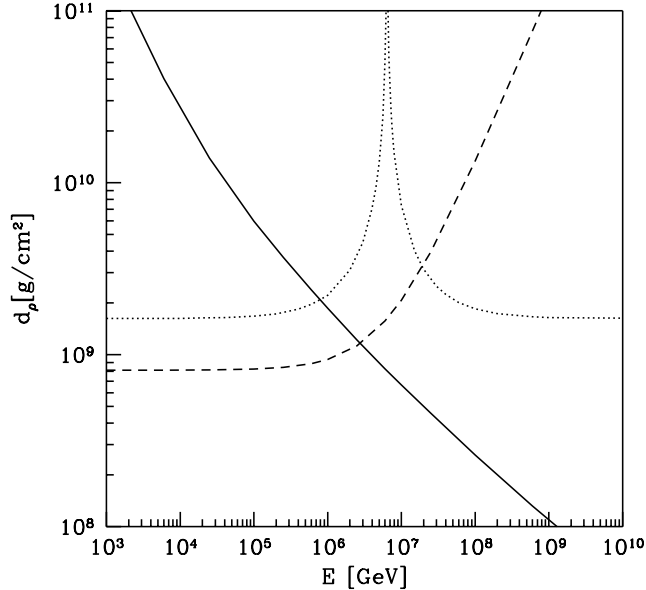, width=9truecm}
\end{center}
\caption{ 
The  width $d_{\rho}=m_N d_{0 N}$ for $\nu_e- \nu_\mu$ (dashed line)
and for $\nu_e- \nu_s$ (dotted line) channels, and
the absorption width, $d_{abs}$, for the electron neutrino (solid line),
as functions of
the neutrino energy. We have considered isotopically neutral medium, $Y_e=0.5$.
The data for $d_{abs}$ are taken from 
ref.\cite{Gandhi:1998ri}.} 
\label{fig:abs} 
\end{figure}

For the active-sterile conversion, one has to take into account also the
neutral current interaction channel ($t$-channel exchange of $Z$), 
for which $q^2=0$, so that the low energy formulas (\ref{eq:pot}-\ref{eq:d0}) are still valid.  
For $\nu_\mu-\nu_s$ only neutral current processes are involved, thus the low 
energy result, eq. (\ref{eq:dstermu}), holds at high energies too. 
In contrast, for the $\nu_e-\nu_s$ case both charged and neutral current interactions contribute,  
and  for an electrically neutral medium the high energy potential can be
written as: 
\begin{equation}
V=\sqrt{2} G_{F}n_e\left({1 \over 1+s_W} -{n_n\over 2 n_p}\right)~.
\label{eq:potsthe} 
\end{equation} 
The second term in eq. (\ref{eq:potsthe}) does not depend on $s_W$, thus coinciding with
the corresponding term in the low energy expression (\ref{eq:potst}).
The potential (\ref{eq:potsthe}) gives the nucleon refraction width:
\begin{equation}
d_{0 N}(s_W)=2 d_0 \left|{1+s_W\over (3Y_e-1)-s_W(1-Y_e)}\right|~.
\label{eq:replasy}
\end{equation}
For isotopically neutral medium ($Y_e=1/2$) we get:
\begin{equation}
d_{0 N}(s_W)=4 d_0 \left|{1+s_W\over 1-s_W}\right|.
\label{eq:replas}
\end{equation}
The width $d_{0 N}(s_W)$ diverges for $s_W\rightarrow 1$ (see fig. \ref{fig:abs}). 
     
At high energies inelastic interactions and absorption become important: at $s_W\sim 1$ the imaginary
part of the interaction amplitude is comparable with the real part.  
In fig. \ref{fig:abs} we show the  refraction width $d_\rho=m_N d_{0 N}$ 
for $\nu_e-\nu_\mu$ and 
$\nu_e-\nu_s$ conversion
and the absorption 
width $d_{abs}$\cite{Gandhi:1998ri,Gandhi:1996tf} as functions of the neutrino energy  $E$
in the rest frame of the matter. We have considered isotopically neutral medium, $Y_e=1/2$.
The absorption width $d_{abs}$ is 
 dominated by the contribution of 
neutrino-nucleon scattering; it decreases monotonically with the energy
$E$. In contrast,  $d_\rho$ starts to increase at $s_W \sim 1$, 
which corresponds to $E=10^6\div 10^7$ GeV, 
according to eq. (\ref{eq:repl}).  For $E \simeq 10^6$ GeV
absorption and refraction become comparable; at higher energies, the 
former effect dominates: $d_{abs}\lta d_0$. This means that for a $\nu_e$  with energy 
$E > 10^6$ GeV the conversion in matter is damped by inelastic interactions and
absorption\cite{Stodolsky:1987dx,Raffelt:1993uj,Sigl:1993fn}, 
therefore a strong conversion effect can not be expected. 

Notice that for small mixing angle $\theta$ the minimum width $d_{min}$
is significantly larger than the refraction width $d_0$, therefore the absorption
starts to be important at lower energies.  Taking, for instance, $\sin 2\theta=0.3$
we have $d_{min}\simeq d_0/\sin 2\theta\simeq 3.3 d_0$, and find that 
$d_{abs}\lta d_{min}$ already for $E\gta 5\cdot 10^{5}~{\rm GeV}$.
 
Let us consider now the matter effect for conversion of antineutrinos. 
For $\bar{\nu}_e- \bar{\nu}_\mu$ channel the only contributing interaction is 
the $\bar{\nu}_e-e$ scattering with $W$ exchanged in the $s$-channel.
In this case $q^2=s$, and using eq. (\ref{eq:ffunc}) we get:
\begin{equation}
d_{0 N}(s_W)={1\over Y_e}{d_0 \over \left|f(s_W)\right|}~.
\label{eq:replantinu}
\end{equation}
This function has a pole at $s_W=1$, i.e., in the $W$-boson resonance. The pole appears because the
amplitude becomes purely imaginary in the resonance, so that the potential is zero.
The width $d_{0 N}(s_W)$ diverges for $s_W\rightarrow \infty$, due to the $1/s_W$ decrease of the
amplitude. The function (\ref{eq:replantinu}) has two minima:
\begin{equation}
d_{0 N}(s^{min}_W)=2\gamma_W {d_0\over Y_e}=2\gamma_W d_{0 N}(s_W=0)
 \hskip 0.9cm {\rm at} \hskip 0.3cm s^{min}_W=1\pm \gamma_W.
\label{eq:minima}
\end{equation}
Numerically $d_{0 N}(s^{min}_W)=0.05~d_{0 N}(s_W=0)$, 
which shows that refraction effects are enhanced close 
to the $W$ resonance. However, in this region inelastic interactions become already important. 

For $\bar{\nu}_e-\bar{\nu}_s$ channel  the contribution of neutrino-nucleon
scattering should be included, and for electrically  neutral medium 
we find:
\begin{equation}
d_{0 N}(s_W)=2 d_0 \left|{(1-s_W)^2+\gamma^2_W\over
{(3Y_e-1)+2s_W(1-2Y_e)-s^2_W(1-Y_e)-\gamma^2_W(1-Y_e)}}\right|~.
\label{eq:d0ye}
\end{equation}
In the case of isotopically neutral matter eq. (\ref{eq:d0ye}) gives:
\begin{equation}
d_{0 N}(s_W)=4 d_0 \left|{(1-s_W)^2+\gamma^2_W\over
{1-s^2_W-\gamma^2_W}}\right|~,
\label{eq:replantinus}
\end{equation}
which has the value $4d_0$ in the limits $s_W\ll 1$ and $s_W\gg 1$, and a pole at 
$s_W\simeq 1$. 
Similarly to eq. (\ref{eq:minima}) we find the minima:
\begin{equation}
d_{0 N}(s^{min}_W)=4\gamma_W d_0=\gamma_W d_{0 N}(s_W=0)
 \hskip 0.9cm {\rm at} \hskip 0.3cm s^{min}_W=1\pm \gamma_W.
\label{eq:minimaas}
\end{equation}

\begin{figure}[hbt]
\begin{center}
\epsfig{file=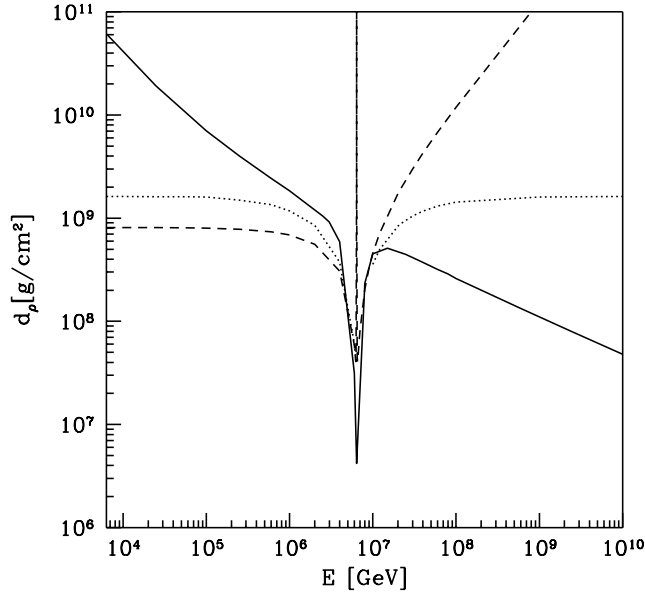, width=9truecm}
\end{center}
\caption{
The width $d_{\rho}=m_N d_{0 N}$ for $\bar{\nu}_e- \bar{\nu}_\mu$ 
(dashed line) and for $\bar{\nu}_e- \bar{\nu}_s$ (dotted line) channels, and
the absorption width, $d_{abs}$, for the electron antineutrino (solid line),
as functions of
the neutrino energy. We have considered isotopically neutral medium, $Y_e=0.5$.
The data for $d_{abs}$  are taken from 
ref.\cite{Gandhi:1996tf}.} 
\label{fig:antinu} 
\end{figure}

In fig. \ref{fig:antinu} we show the  refraction width $d_\rho=m_N d_{0 N}$ for
$\bar{\nu}_e-\bar{\nu}_\mu$ and $\bar{\nu}_e-\bar{\nu}_s$ channels
and the absorption width for the electron antineutrino, 
$d_{abs}$\cite{Gandhi:1996tf}, as functions of the neutrino energy. We have considered isotopically neutral medium.
 For energies outside the $W$ boson resonance interval the
main contribution to $d_{abs}$ is given by the neutrino-nucleon scattering; 
at $s_W\simeq 1$ the effect of the resonant $\bar{\nu}_e-e$ scattering dominates,
providing the narrow peak.    
It appears that absorption prevails over refraction ($d_{abs}<d_0$)
for $E\gta 6\cdot 10^{6}\thinspace 
{\rm GeV}$, corresponding to $d_\rho\simeq 6\cdot 10^{7} {\rm g\cdot cm^{-2}}$, for both 
$\bar{\nu}_e-\bar{\nu}_\mu$ and $\bar{\nu}_e-\bar{\nu}_s$ cases.
 
The effect of absorption on neutrino conversion
starts to be important at lower energies: 
for  $\sin 2\theta=0.3$
 we find that  $d_{abs}\lta d_{min}$ at  
$E\gta 6\cdot 10^{5}~{\rm GeV}$.


\subsection{High energy neutrinos in neutrino environment}

Let us consider a beam of neutrinos which propagates in a background made of neutrinos of very low
energies\footnote{We will not consider the conversion of neutrinos in the background itself,
which can significantly affect the flavour content of the background.}. 
This could be the case of beams of low energy neutrinos from supernovae, or high
energy neutrinos from AGN and GRBs, or neutrinos produced by the annihilation of superheavy 
relics, etc..  We assume that the background consists of neutrinos and antineutrinos of various
flavours, with number densities $n_i$ ($i=\nu_e, \bar{\nu}_e, \nu_\mu, $ etc.). In the case of
relativistic neutrino background we assume its isotropy.

The potential for a neutrino $\nu_{\alpha}$ ($\alpha=e,\mu,\tau$)
due to neutrino-neutrino interaction can be written as:
\begin{eqnarray}
&&V_{\nu_\alpha}(s_Z)= \sqrt{2} G_{F}\left[  
n_{\nu_{\alpha}}f(-s_Z)
-n_{\bar{\nu}_{\alpha}} f(s_Z) 
+\sum_{i=\nu_e,\nu_\mu,\nu_\tau}\left( n_i-n_{\bar{i}}\right) \right]~,
\label{eq:potnumu} 
\end{eqnarray} 
where the propagator function $f(s_Z)$ has been defined in eq. (\ref{eq:ffunc}).  
Here $s_Z\equiv s/M^2_Z$ and $\gamma_Z\equiv \Gamma_Z/M_Z$;
$M_Z$ and $\Gamma_Z$ are the mass
and width of the $Z$-boson.  
The first term in eq. (\ref{eq:potnumu})  is due to
 $\nu_{\alpha}-\nu_{\alpha}$ scattering with $Z$-boson exchange in $u$-channel, and the second term is
the contribution from $\nu_{\alpha}-\bar\nu_{\alpha}$ annihilation. 
\\


\subsection{Neutrino conversion in CP-asymmetric neutrino background}
As a first case we consider a strongly CP-asymmetric neutrino background, and 
suppose $n_i\gg n_{\bar i}$, so that we can neglect the contributions 
of antineutrinos in (\ref{eq:potnumu}).  
For simplicity, we assume equal concentrations
for the neutrino species: $n_{\nu_{e}}=n_{\nu_{\mu}}=n_{\nu_{\tau}}$.
In terms of the total number density of neutrinos,  
$n_\nu\equiv n_{\nu_{e}}+n_{\nu_{\mu}}+n_{\nu_{\tau}}$, the potential 
(\ref{eq:potnumu}) reduces to:
\begin{eqnarray}
V_{\nu_\alpha}(s_Z)&=&\sqrt{2} G_{F} n_{\nu}\left[1+ {1\over 3}f(-s_Z) \right]~. 
\label{eq:potnumuasym} 
\end{eqnarray}
The potential for the antineutrino is given by $V_{\bar{\nu}_\alpha}(s_Z)=-V_{\nu_\alpha}(-s_Z)$.
\\

Let us now find the refraction width $d_0$ and $d_{min}$ for various channels. 

\begin{figure}[hbt]
\begin{center}
\epsfig{file=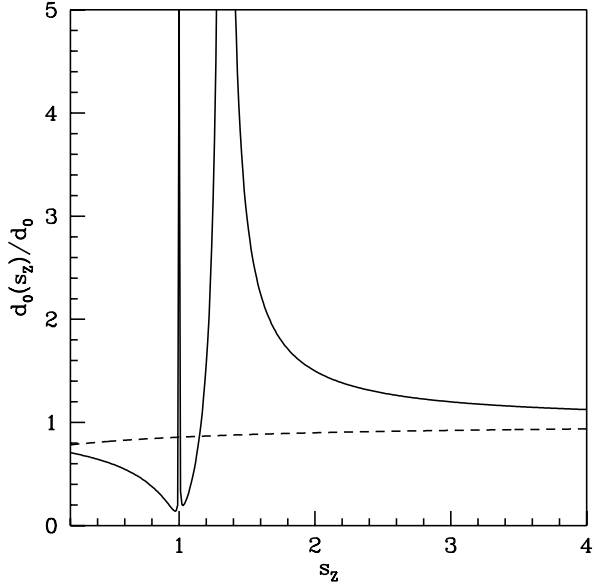, width=9truecm}
\end{center}
\caption{The dependence of the refraction width $d_0$ for 
$\nu_\alpha - \nu_s$ (dashed line) and $\bar{\nu}_\alpha - \bar{\nu}_s$ (solid line)
channels on $s_Z$ in strongly CP-asymmetric
background, $n_i\gg n_{\bar i}$. Equal concentrations are assumed for the various flavours.
} 
\label{fig:A} 
\end{figure}

1). For the active-sterile conversion, $\nu_\alpha - \nu_s$, the potential (\ref{eq:potnumuasym})
coincides with the 
difference of the potentials for the two species, and therefore, by eq. 
(\ref{eq:d0def}), it 
gives immediately the refraction width of neutrinos:
\begin{equation}
d_0(s_Z)=d_0 \left|1+ {1\over 3}f(-s_Z)\right|^{-1}~.  
\label{eq:dun}
\end{equation}
The width $d_0(s_Z)$ is constant for $s_Z\ll 1$ and $s_Z\gg 1$: 
$d_0(s_Z\ll 1)\simeq 3d_0/4=1.84\cdot 10^{32} {\rm cm^{-2}}$, and $d_0(s_Z\gg 1)\simeq d_0$ (see
fig.\ref{fig:A}).

Let us now compare the refraction and absorption effects. 
The main contribution to the absorption width $d_{abs}$\cite{Roulet:1993pz} is
given by the $\nu_\alpha-\nu_\alpha$ and $\nu_\alpha-\nu_\beta$ ($\beta\neq \alpha$)
scatterings.  The width $d_{abs}$ decreases monotonically with 
$s_Z$ and at $s_Z\gg 1$ it takes the value $d_{abs}(s_Z\gg 1)\simeq \pi/(2 G^2_F M^2_Z)\simeq 3.6\cdot
10^{33}~{\rm cm^{-2}}$.
Due to its non-resonant behaviour, $d_{abs}$ is
larger than $d_0$ for any energy of the neutrinos: at $s_Z\gg 1$ we find that 
$d_0/d_{abs}\simeq G_F M^2_Z/\sqrt{2}=\pi \alpha_W/(2 \cos^2 \theta_W) \simeq 0.1$, where 
$\theta_W$ and $\alpha_W=g^2/4\pi$ are the weak mixing angle and coupling constant.   
Therefore, $d_{abs}$  is also larger than the minimum width, 
$d_{min}$, for  $\sin 2\theta\gta d_0/d_{abs}\simeq 0.1$. 
\\

2). For the conversion of an active antineutrino into a sterile species, $\bar{\nu}_\alpha-\bar{\nu}_s$, 
we get the width:
\begin{equation}
d_0(s_Z)=d_0 \left|1+ {1\over 3}f(s_Z)\right|^{-1}~.  
\label{eq:dunanti}
\end{equation}
This function (see fig.\ref{fig:A}) has a resonant behaviour with minima 
at $s_Z\simeq 1\pm \gamma_Z$: $d_0(1-\gamma_Z)\simeq (1/6\gamma_Z+1)^{-1}d_0\simeq d_0/7$ and 
$d_0(1+\gamma_Z)\simeq (1/6\gamma_Z-1)^{-1}d_0\simeq d_0/5$.
Outside the $Z$-boson resonance $d_0(s_Z)$ is constant: 
$d_0(s_Z\ll 1)\simeq 3d_0/4$ and $d_0(s_Z\gg 1)\simeq d_0$. 
In the range $s_Z\sim 1$ inelastic
scattering and absorption become important.
We evaluate the absorption width $d_{abs}$ for antineutrino in neutrino background using the
plots in ref.\cite{Roulet:1993pz}.  
For $s_Z<1$, the width $d_{abs}$ decreases with the increasing $s_Z$; at $s_Z\simeq 1$ it shows the
characteristic peak due to  the resonant $\nu_\alpha-\bar{\nu}_\alpha$ scattering.
For $s_Z\gta 1$ the absorption width increases with $s_Z$ up to the limit 
$d_{abs}(s_Z\gg 1)\sim 3\cdot 10^{33}~{\rm cm^{-2}}$, due to the contributions of
$\nu_\beta-\bar{\nu}_\alpha$ scatterings ($\beta\neq \alpha$) and 
$\nu_\alpha-\bar{\nu}_\alpha$ interaction in the $t$-channel.
We find that $d_{abs}\gta d_0(s_Z)$ for $s_Z\lta 0.8$ ($s\lta 7\cdot 10^{3} {\rm GeV^2}$) and for 
$s_Z\gta 2$ ($s \gta 1.7\cdot 10^{4} {\rm GeV^2}$). 
Furthermore, $d_0(s_Z=0.8)\simeq 10^{32} {\rm cm^{-2}}$ and 
$d_0(s_Z=2)\simeq 4\cdot 10^{32} {\rm cm^{-2}}$.
Taking $\sin 2\theta =0.3$ we get that $d_{abs}\gta d_{min}$
for $s_Z\lta 0.7$  and $s_Z\gta 2.4$; 
$d_{min}(s_Z=0.7)\simeq 3\cdot 10^{32} {\rm cm^{-2}}$ and 
$d_{min}(s_Z=2.4)\simeq 10^{33} {\rm cm^{-2}}$.

Notice that, in contrast with the 
conversion in matter (see figs.\ref{fig:abs}  and \ref{fig:antinu}), for 
neutrinos and antineutrinos in
neutrino environment we can have $d_{abs}\gta d_{min}(s_Z)$ even in the high energy range, $s_Z\gta 1$:
in particular, we find that $d_{abs}(s_Z\gg 1)\gta d_{min}(s_Z\gg 1)$ for $\sin 2\theta\gta 0.1$. 
\\ 

3). Let us now consider the active-active conversion, $\nu_\alpha - \nu_\beta$.
Assuming equal concentrations of neutrinos of different flavours, 
$n_{\nu_{e}}=n_{\nu_{\mu}}=n_{\nu_{\tau}}$, we find
from eq. (\ref{eq:potnumuasym})  that the difference of the potentials of the
two species equals:
\begin{eqnarray}
\Delta V_{\alpha,\beta}=V_{\nu_\alpha}-V_{\nu_\beta}
={1\over 3}\sqrt{2} G_{F} n_{\nu}\left[f(-s^{\alpha}_Z)-f(-s^{\beta}_Z) \right]~, 
\label{eq:potabasym} 
\end{eqnarray} 
where  $s^{i}_Z\equiv s^i/M^2_Z$ ($i=\alpha, \beta$), and $s^{i}$ is the center of mass 
energy squared of the incoming and the background neutrino of the same type $i$. 
If the  species $\nu_\alpha$ and $\nu_\beta$ in the background have different energies we get that
$s^{\alpha}_Z\neq s^{\beta}_Z$, and therefore $\Delta V_{\alpha,\beta}\neq 0$, leading to matter
induced neutrino conversion even if $\nu_\alpha$ and $\nu_\beta$ have equal concentrations.
This situation is realized if the background neutrinos $\nu_\alpha$ and $\nu_\beta$ have different
masses, e.g. $m_{\nu_\alpha}>m_{\nu_\beta}$, and are
non-relativistic.
Denoting by $E$ the energy of the neutrino beam, in the rest frame of the background we have 
$s^{i}=2m_i E$, and thus  $s^{\alpha}_Z/s^{\beta}_Z= m_{\nu_\alpha}/m_{\nu_\beta}> 1$.
The condition $s^{\alpha}_Z\neq s^{\beta}_Z$ is achieved also 
if one of the neutrino species is
relativistic and the other is not: $m_{\nu_\alpha}\gg E_\beta \gg m_{\nu_\beta}$, 
where $E_\beta$ is the energy of $\nu_\beta$ in the
background. 
Assuming the isotropy of the neutrino gas, we have that   
$s^{\beta}\simeq 2 E_\beta E$.  

\begin{figure}[hbt]
\begin{center}
\epsfig{file=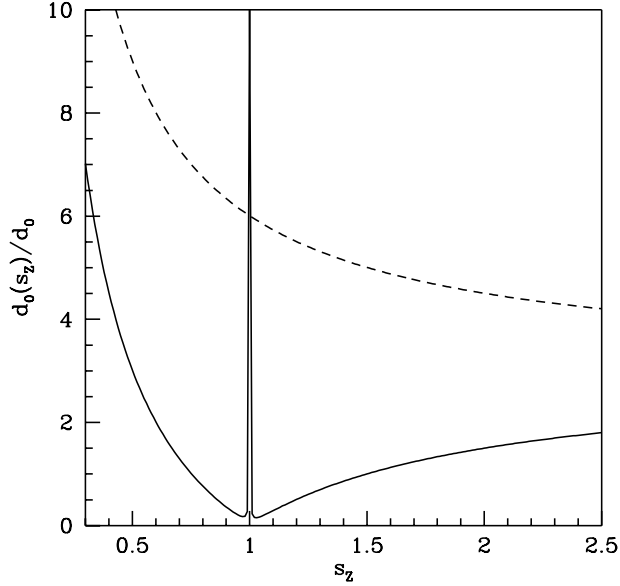, width=9truecm}
\end{center}
\caption{The dependence of the refraction width $d_0$ for 
$\nu_\alpha - \nu_\beta$ (dashed line) and $\bar{\nu}_\alpha - \bar{\nu}_\beta$ (solid line)
channels on $s_Z$ in strongly 
CP-asymmetric
background, $n_i\gg n_{\bar i}$. Equal concentrations are assumed for the various flavours.
We have considered $s^\alpha_Z\gg s^\beta_Z\sim 0$. } 
\label{fig:B} 
\end{figure}

Using (\ref{eq:potabasym}) and (\ref{eq:d0def}), we get the refraction width:
\begin{equation}
d_0(s^{i}_Z)=3 d_0 \left| f(-s^{\alpha}_Z)-f(-s^{\beta}_Z)\right|^{-1} 
\simeq 3 d_0 \left|(1+s^\alpha_Z)(1+s^\beta_Z) \over (s^\alpha_Z-s^\beta_Z)\right|~. 
\label{eq:dunab}
\end{equation}
The function (\ref{eq:dunab}) diverges for $s^{i}_Z\rightarrow \infty$ and $s^{i}_Z\rightarrow 0$.
In particular, in the low energy limit, $s_Z\ll 1$, it reduces to 
$d_0(s^{i}_Z\ll 1)\simeq 3 d_0/(2 E \Delta m)$, where $\Delta m\equiv m_{\nu_\alpha}-m_{\nu_\beta}$. 
For the realistic case $s^{\alpha}_Z\gta 1$ and $s^{\beta}_Z\ll 1$, eq. (\ref{eq:dunab}) can be
written as:
\begin{equation}
d_0(s^{\alpha}_Z)
\simeq 3 d_0 \left|{1+s^\alpha_Z \over s^\alpha_Z }\right|~, 
\label{eq:dunabapp}
\end{equation}
which approaches the minimum value $3d_0$ when  $s^{\alpha}_Z\gg 1$ (see fig.\ref{fig:B}).
Taking the maximal realistic values for the mass and energy of the neutrino,
$m_{\nu_\alpha}=5$ eV and  $E=10^{22}$ eV we get $s^\alpha_Z\simeq 12$ at most, so that  
$d_0(s^{i}_Z )\simeq 3.5 d_0$.
\\

4). For the $\bar{\nu}_\alpha - \bar{\nu}_\beta$ channel the effective potential equals:
\begin{eqnarray}
\Delta V_{\alpha,\beta}
={1\over 3}\sqrt{2} G_{F} n_{\nu}\left[f(s^{\alpha}_Z)-f(s^{\beta}_Z) \right]~, 
\label{eq:potabasymCP} 
\end{eqnarray}
and therefore we get the width:  
\begin{eqnarray}
d_0(s^{i}_Z)
=3 d_0 \left| f(s^{\alpha}_Z)-f(s^{\beta}_Z)\right|^{-1}~. 
\label{eq:dunabCP}
\end{eqnarray}
Due to the resonant character of the function $f(s_Z)$, the width $d_0(s^{i}_Z)$ has the following
features (see fig.\ref{fig:B}):

\noindent
(i) It reaches the  local minimum $d_0(s^{i}_Z)\simeq 6 \gamma_Z d_0\sim d_0/6$ when one of the 
$s^i_Z$'s is at resonance and the other is outside the resonance: 
$s^{\alpha}_Z\simeq 1 $ and $s^{\beta}_Z\neq 1$ (or vice versa).

\noindent
(ii) The absolute minimum $d_0(s^{i}_Z)\simeq 3 \gamma_Z d_0$ is achieved when
$s^{\alpha}_Z\simeq 1+\gamma_Z$ and $s^{\beta}_Z\simeq 1-\gamma_Z$ (or vice versa). These conditions
can be satisfied for certain relations between the masses of the background neutrinos.
For non-relativistic background: $m_{\nu_\alpha}/m_{\nu_\beta}=(1+\gamma_Z)/(1-\gamma_Z)$.

Notice that for $s^{i}_Z$ discussed in (i) and (ii) the effects 
of inelastic scattering and absorption can
be important.

\noindent 
(iii) If one of the $s^i_Z$'s is far below the resonance and the other is far above 
(e.g. $s^{\alpha}_Z\gg 1 $ and $s^{\beta}_Z\ll 1$) then 
$d_0(s^{i}_Z)\sim 3 d_0$. 

\noindent
(iv) $d_0(s^{i}_Z)\gg d_0$ if both the $s^i_Z$'s are far below or far above the resonance. 

Obviously, for strong CP-asymmetric background with $n_{\bar i}\gg n_i$ the results for $\nu$ and
$\bar{\nu}$ channels should be interchanged.
\\

\subsection{Neutrino conversion in CP-symmetric neutrino background}
Let us now consider the neutrino conversion in a CP-symmetric neutrino background, $n_i=n_{\bar i}$, 
with $n_{\nu_{e}}=n_{\nu_{\mu}}=n_{\nu_{\tau}}$.
In this case the potential (\ref{eq:potnumu}) can be written as:
\begin{eqnarray}
V_{\nu_\alpha}
&=&\sqrt{2} G_{F} n_{\nu_{\alpha}} \left[f(-s_Z)- f(s_Z)\right]~.  
\label{eq:potnumusym} 
\end{eqnarray}  
It vanishes in the low energy limit $s_Z\rightarrow 0$, but it is unsuppressed at high energies, 
$s_Z\gta 1$, leading to significant matter effect.  We will consider the conversion of neutrinos; 
due to the CP-symmetry antineutrinos undergo analogous effects.

\begin{figure}[hbt]
\begin{center}
\epsfig{file=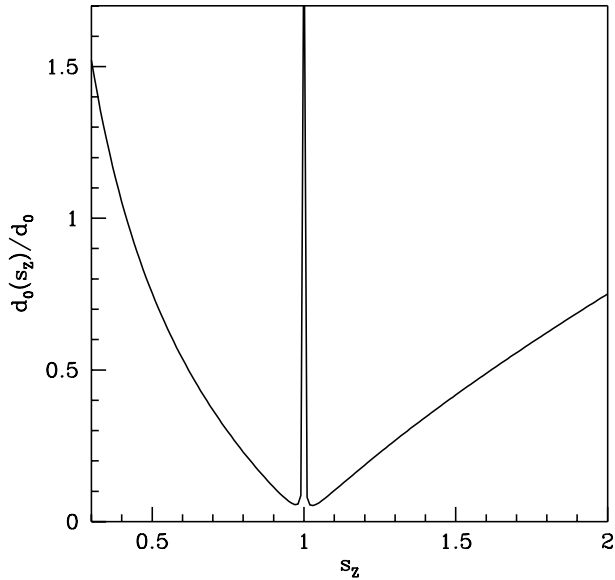, width=9truecm}
\end{center}
\caption{The dependence of the refraction width $d_0$ for 
$\nu_\alpha - \nu_s$ and $\bar{\nu}_\alpha - \bar{\nu}_s$ channels on $s_Z$  
in CP-symmetric neutrino
background. } 
\label{fig:C} 
\end{figure}

1). For $\nu_\alpha - \nu_s$ conversion from the potential (\ref{eq:potnumusym}) we find 
the refraction width for the neutrinos of flavour $\alpha$:
\begin{eqnarray}
d_0(s_Z)&=&d_0 \left|f(-s_Z)-f(s_Z)\right|^{-1}  \nonumber \\
&=&d_0 \left| {(1+s^2_Z+\gamma^2_Z)^2-4s^2_Z \over 2 s_Z [(1-s^2_Z)+ \gamma^2_Z]}\right|~.
\label{eq:dhalo}
\end{eqnarray}
For $s_Z\ll 1$ it behaves as $d_0(s_Z)\simeq d_0/(2 s_Z)$ and for 
$s_Z\gg 1$ we have $d_0(s_Z)\simeq d_0 s_Z/2$ (see fig.\ref{fig:C}).
The width (\ref{eq:dhalo}) has two minima:
\begin{equation}
d_0(s^{min}_Z)\simeq 2\gamma_Z d_0  \hskip 0.9cm {\rm at} \hskip 0.3cm s^{min}_Z=1\pm \gamma_Z~.
\label{eq:minimann}
\end{equation}
Numerically, $d_0(s^{min}_Z)=0.055~d_0\simeq 1.35 \cdot 10^{31} {\rm cm^{-2}}$.

The absorption width $d_{abs}$ is dominated by  
$\nu_\alpha-\bar{\nu}_\alpha$ annihilation, with a resonance peak 
at $s_Z\sim 1$. 
Using the results of ref.\cite{Roulet:1993pz} we find that
 $d_{abs}$ is larger than $d_0(s_Z)$ outside the $Z$-boson resonance, and 
the two quantities are comparable
at $s_Z\sim 1$ or at $s_Z\gg 1$.
For $\sin 2\theta=0.3$ we find that the minimum width 
$d_{min}$ is larger than $d_{abs}$ in the range $ 0.7\lta s_Z \lta 1.6$, corresponding to 
$6\cdot 10^{3} {\rm GeV^2} \lta s \lta 1.3\cdot 10^{4} {\rm GeV^2}$.  At the edges of this interval the
width $d_{min}$ takes the value 
$d_{min}(s_Z=0.7)\simeq d_{min}(s_Z=1.6)\simeq 3 \cdot 10^{32} {\rm cm^{-2}}$.
\\

2). For the $\nu_\alpha - \nu_\beta$ channel, 
eq. (\ref{eq:potnumusym}) gives the difference of potentials:
\begin{eqnarray}
\Delta V_{\alpha,\beta}&=&V_{\nu_\alpha}-V_{\nu_\beta} \nonumber \\
&=&\sqrt{2} G_{F} n_{\nu_\alpha}\{[f(-s^{\alpha}_Z)-f(s^{\alpha}_Z)]
-[f(-s^{\beta}_Z)-f(s^{\beta}_Z)]\}~.
\label{eq:potabsym} 
\end{eqnarray}
The corresponding refraction width equals:
\begin{equation}
d_0(s^{i}_Z)=d_0\left|[f(-s^{\alpha}_Z)-f(s^{\alpha}_Z)]
-[f(-s^{\beta}_Z)-f(s^{\beta}_Z)] \right|^{-1}~. 
\label{eq:dabcpsym}
\end{equation} 
We find that 
$d_0(s^{i}_Z)\gta d_0$ when both $s^{\alpha}_Z$ and $s^{\beta}_Z$  are outside the $Z$-boson resonance
and $d_0(s^{i}_Z)$ takes its minimum values when either 
$s^{\alpha}_Z$ or $s^{\beta}_Z$ is close to the $Z$-resonance:

If $s^{\alpha}_Z\sim 1+\gamma_Z$ and $s^{\beta}_Z\sim 1-\gamma_Z$ (or vice versa)
$d_0(s^{i}_Z)$ has the absolute minimum 
$d_0(s^{i}_Z)\simeq \gamma_Z d_0$. 
For $s^{\alpha}_Z\sim 1 $ and $s^{\beta}_Z\neq 1$ (or vice versa) we have the local minimum 
$d_0(s^{i}_Z)\simeq 2 \gamma_Z d_0$. 
Notice that in the realistic case  $s^{\alpha}_Z\gg s^{\beta}_Z\sim 0$ the width (\ref{eq:dabcpsym})
reduces essentially to the one in eq. (\ref{eq:dhalo}).
At resonance, where $d_0(s^{i}_Z)$ has minima, the effects 
of inelastic collisions and absorption are important.   



\section{Applications}
The results derived in the previous sections are now applied to some physical
situations of interest.  After a brief discussion of well known cases, like the
Earth, the Sun and supernovae, we present results for neutrinos in some new
astrophysical environments. We find that significant matter induced conversion
can be expected for neutrinos crossing the dark matter halos of clusters of galaxies 
and for neutrinos from cosmologically distant sources.


\subsection{Minimum width condition and bounds on the mixing}
As follows from the analysis in section 2, a significant neutrino conversion in matter
requires the fulfilment of the minimum width condition\footnote{This condition
refers to the requirement of conversion probability larger than $1/2$, eq. (\ref{eq:half}).
In some circumstances, however, 
even a small effect, with conversion probability $P\ll 1/2$ can be important.}:
\begin{equation} 
d\ge d_{min}={d_0\over \tan 2\theta}~.
\label{eq:ineq}
\end{equation}
This condition is independent of the density distribution, and therefore of the specific
matter effect involved. Thus the knowledge of the width $d$ allows one to conclude about
the significance of the matter effect even if the density profile is unknown. This is the
case of some astrophysical objects for which 
estimates or bounds on $d$ can be
obtained directly by observational data with no assumption on their internal structure.

In the Table \ref{tab:tab1} we show the parameters of interest of some objects, 
toghether with the values of the ratio
\begin{equation} 
r\equiv {d\over d_0}~.
\label{eq:defr}
\end{equation}
For $r<1$, and small mixing angle, the condition (\ref{eq:ineq}) can not be satisfied, 
thus no significant neutrino conversion is expected.  Conversely, for $r\gta 1$,   
(\ref{eq:ineq}) can be fulfilled and gives the bound on the mixing:
\begin{equation} 
\sin 2\theta \gta {1\over r}={d_0\over d}~. 
\label{eq:boundsin}
\end{equation}
Notice that our analysis holds for small mixings: $\sin 2\theta\ll 1$. For applications we assume
$\sin 2\theta\lta 0.3$, for which we find from eq. (\ref{eq:boundsin}) that the minimum width
condition is satisfied for 
$r\gta 3$.

The inequality (\ref{eq:boundsin}) can be considered as the sensitivity limit for the mixing angle
that can be achieved by studies of neutrino conversion in a layer of  given width $d$.
The real sensitivity can be however much lower than the absolute limit given by the condition 
(\ref{eq:boundsin}).  This is related to the fact that in the case of varying density only part of 
the total amount of matter effectively contributes to the conversion.
Introducing the corresponding width $d^{conv}$ we have the condition:
\begin{equation} 
\sin 2\theta \gta {d_0\over d^{conv}}~, 
\label{eq:boundsineff}
\end{equation}
instead of the (\ref{eq:boundsin}).

Let us find the expression of $d^{conv}$ for a medium with monotonically varying density.
As discussed in section 2.3, the transition occurs mainly in the resonance layer.  Using the result 
(\ref{eq:deltan}) we get:
\begin{equation} 
d^{conv}=n^{res}_e{dL\over dn} 2\Delta n={2\over \sqrt{3}}n^{res}_e l_n
\tan2\theta~,
\label{eq:dconv}
\end{equation} 
where $l_n\equiv |({dn\over dL})^{-1}|_{res}n^{res}_e$.
Inserting the expression (\ref{eq:dconv}) in the condition (\ref{eq:boundsineff}), we find:
\begin{equation} 
\sin^2 2\theta \gta {\sqrt{3} d_0\over 2 n^{res}_e l_n}~. 
\label{eq:bound2}
\end{equation} 
Clearly, $d^{conv}$ could be much smaller than the total width $d$ of the object, 
so that the condition (\ref{eq:bound2}) on the mixing could be much stronger than (\ref{eq:boundsin}).
Notice that the bound (\ref{eq:bound2}) is quadratic in $\sin 2\theta$.
Using the definition (\ref{eq:adiab}) of the adiabaticity parameter, $\gamma$,  
the condition (\ref{eq:bound2}) can be written as $\gamma\leq 4/(\pi\sqrt{3})$,
which corresponds to the adiabaticity condition close to its limit of validity. 

Another important issue is that the maximal sensitivity for the mixing $\theta$ can be achieved for
particular values of $\Delta m^2/E$, which depend on the specific density profile.
As follows from (\ref{eq:bound2}), for constant (or slowly varying with the distance) $l_n$ 
the smallest $\sin^2 2\theta $
corresponds to the largest $n^{res}_e$, and therefore to the largest values 
of $\Delta m^2/E$. This is the case of exponential density profile. For power-law profile,
$n_e\sim L^{-k}$, we get $|l_n|={L/ k}$, so that $\sin^2 2\theta\sim L^{k-1}$. Taking $k>1$, fulfilled
by practically all the realistic profiles, we find that the smallest $\theta$ is achieved for the
smallest $L$, and consequently the highest values of $n_e$ and $\Delta m^2/E$.

Notice that $d^{conv}$ is a local property which depends on the derivative in $l_n$. Of course, the
description given by $d^{conv}$ is not correct when the density profile is close to the constant
one, so that $l_n\rightarrow \infty$. In this case $d^{conv}$ can be even larger than the total
width $d$. 
Thus, the correct condition on the mixing can be written as:
\begin{equation}
\sin 2\theta\gta {d_0 \over min\left[d, d^{conv}\right]}~.
\label{truecond}
\end{equation} 

\begin{table}[hbt]
\centering
\begin{tabular}{|c|c|c|c|}
\hline
object & density (${\rm cm^{-3}}$) & size (cm) &  
$r=d/ d_{0}$ \\
\hline
 Earth:  & & & \\
 $\cos\theta_z=1$  &   $2.6\cdot10^{24}$     &  $ 1.26 \cdot 10^9 $    &      
 $ 13.6 $ 
           \\
$\cos\theta_z=0.81$ & $ 1.5\cdot 10^{24}$    &   $ 10^9$ &  $6.4 $    \\
\cline{2-3}
\hline

Sun  & $\sim 7\cdot 10^{24}$     & $6.96 \cdot 10^{10} $& $2600 $ \\
\hline
                                                                  
Moon   &$ \sim 10^{24}$  &  $3.48 \cdot 10^8$  & $1.4 $\\
\hline
                                                                                                                                    
Supernova   & $3\cdot 10^{33}$   & $10^7 $& $10^9  $\\
\hline

Universe & & & \\ 
($n_\nu=n_{\bar{\nu}}$)   & $ 1.5\cdot 10^{4}  $  & $10^{27}$ & $ 3\cdot 10^{-2} $\\ 
\hline

Universe & & & \\ 
($n_\nu\gg n_{\bar{\nu}}$)   & $ \sim 10^{5}  $  & $10^{27}$ & $ 0.3 $\\ 
\hline

Galactic halo  & $\lta 2 \cdot 10^6 $ &
 $3 \cdot 10^{23}$ & $ 5\cdot 10^{-2} $\\

\hline

Cluster halo  & $\lta 5\cdot 10^7 $ &
 $3 \cdot 10^{24}$ & $ 10  $\\

\hline
                                                                  
AGN   & & $d \simeq  10^{22} \div 10^{23} {\rm  cm^{-2}}$ & $10^{-10} \div 10^{-9}$  \\
\hline
                                                                  
GRB   & $10^{10} \div 10^{12}$  & $< 5 \cdot 10^{15}$ &  $< 10^{-5}$\\
\hline
\end{tabular}
\caption{The density, the size and the matter width in units of refraction width,
$r=d/d_{0}$, for various physical objects.
The values given for the densities are averaged along the
trajectories of the neutrinos. We quote the number density of electrons for  
objects made of usual matter,
and the concentration of the neutrino background for the halos and the universe.
For the Earth the results are given for two trajectories
with different zenith angle $\theta_z$.  The results for the universe correspond to 
redshift $z=5$
for the cases of $\nu_\alpha - \nu_s$ and  $\bar{\nu}_\alpha-\bar{\nu}_s$ in CP-symmetric and strongly
CP-asymmetric  neutrino background with $\eta_\nu\simeq 1$.
}
\label{tab:tab1}
\end{table}  
                                   

\subsection{The Sun, the Earth, the Moon and supernovae}
For neutrinos crossing the Earth we consider two types of trajectories,
 corresponding to different values of the zenith angle $\theta_z$.
For $\cos \theta_z$=1 neutrinos travel along the diameter of the Earth,
crossing the core and the two layers of the mantle.  We get  $r=$13.6, and therefore 
according to (\ref{eq:boundsin}) we could expect significant matter 
conversion for $\sin^2 2\theta \gta 5\cdot 10^{-3}$. 
However this maximal sensitivity, which would be achieved for uniform density distribution, 
is not realized for the Earth profile. For small mixing, the difference between the densities in the
core and in the mantle is larger than the resonance interval.  As a result, the oscillations are
resonantly enhanced either in the mantle or in the core, and only one of the two parts effectively
contributes to the effect.  At the same time, for certain ranges of $\Delta m^2/E$, different from
both the resonance values in the core and in the mantle, parametric enhancement of oscillations
occurs.  Numerical calculations\cite{MikSm} give $\sin^2 2\theta \gta 2\cdot 10^{-2}$ as best sensitivity. 

For  $\cos \theta_z$=0.81 the trajectory is tangential to the core, and therefore it represents
the path  of maximal length in
the mantle.  In this case we find $r\simeq$6.4 and the sensitivity limit 
$\sin^2 2\theta \gta 2.5\cdot 10^{-2}$.  Since this case realizes approximatively
the optimal condition of uniform medium, we have good agreement with the results of exact
calculations. 

In the case of the Moon, $r=$1.4, and therefore a large mixing is required:
$\sin^2 2\theta \gta $0.5.

A numerical integration of the density profile of the Sun\cite{bah}
gives $d\simeq 1.5\cdot 10^{12} {\rm g\cdot cm^{-2}}$. Dividing this result by $d_\rho=m_N d_{0 N}$,
with $Y_e=0.7$, we find $r\simeq$2600. 
From the condition (\ref{eq:boundsin}) we get then $\sin^2 2\theta \gta 1.5\cdot10^{-7}$.  
This bound is remarkably
weaker than the one obtained from the condition (\ref{eq:bound2}): 
taking $n^{res}_e\simeq 50 A 
\thinspace {\rm cm^{-3}}$ and $l_n\simeq 0.3 R_\odot$, we get 
$\sin^2 2\theta\simeq 2.4\cdot10^{-4}$, in good agreement with the results 
of exact computations\cite{Mikheev:1986wj,MikSm}. 

For supernovae the total width of the matter above the neutrinosphere gives 
 $r\simeq 10^9$, for which the condition (\ref{eq:boundsin}) would lead to 
 $\sin^2 2\theta \gta 10^{-18}$.
Using  the
density profile $n_e=n^0_e (R_0/R)^{3}$\cite{Notzold:1987qt}, with $R_0=10^{7}$ cm and 
$n^0_e\simeq 10^{34} {\rm cm^{-3}}$, from (\ref{eq:bound2})
we find $\sin^2 \theta\gta 10^{-8}$, which
agrees well with the results of numerical calculations\cite{MikSm}. 

As shown in the previous examples, the maximal sensitivity for $\sin^2 2\theta$, given by the total
width $d$, can be achieved in the case of uniform medium at $\Delta m^2/E$ corresponding to the
resonance density.  Such a situation is realized for neutrinos crossing the mantle of the Earth.
In the case of substantial deviations from the constant density, like in the Sun or in supernovae, 
the sensitivity is much lower.  The stronger the deviation  from constant density, 
the smaller $d^{conv}$, and therefore the
lower is the sensitivity.


\subsection{AGN and GRBs}
Let us now turn to high energy 
neutrinos from Active Galactic Nuclei (AGN) and Gamma Ray Bursters 
(GRBs)\cite{Protheroe:1998dm, Ginzburg:1990sk,Blinnikov:1999wz,Piran:1999bk}.

In AGN, neutrinos are considered to be produced by the interaction of
accelerated protons with a photon or proton 
background\cite{Stecker:1991vm,Szabo:1994qx,Nellen:1993dw}. 
There is a hope that neutrinos from AGN with energies $\gta 10^6$ GeV could be detected by large
scale underwater (ice) and EAS detectors\cite{Halzen:1998mb}.

The width of matter crossed by neutrinos in an AGN can be estimated 
on the basis of the 
existing data on the X-ray emission of these objects.  The variability of the
spectra suggests that the X-radiation is emitted very
close to the AGN core\cite{Ginzburg:1990sk}.
The proton acceleration and therefore the neutrino production are supposed to happen in the same
region. For this reason the width of matter crossed by neutrinos equals approximatively the one
crossed by the X-radiation.  For the later  
the experimental data\cite{Mushotzky:1982bh, 
Turner:1989jq,Sambruna:1999qe} give the value 
$d_{AGN} \simeq (10^{-2} \div 10^{-1})~A$ ${\rm cm^{-2}}$, therefore 
 significant neutrino conversion in AGN is 
excluded\footnote{In the present discussion we have considered radial propagation of neutrinos from the
inner to the external regions of the object. We have not considered  neutrinos travelling through the
core of the AGN.  In this case a significant matter-induced conversion could occur, however neutrinos
crossing the core are supposed to be a small fraction of the total neutrino flux produced.}
 (for a short discussion, see also ref.\cite{Minakata:1996nd}). 

A rather successful description of the origin of GRB is provided by
the fireball model\cite{rees}, in which neutrino production is predicted to happen in an
analogous way as in 
AGN\cite{Waxman:1997hj,Waxman:1999ai }.  
A fireball can emit protons, detected as
high-energy cosmic rays on Earth, accompanied by a flux of neutrinos.
The requirement that the fireball should be transparent to protons gives an
estimate of the width of the object: $d_{GRB} \leq d_{abs}$, where
$d_{abs}=10\div 100~A\thinspace {\rm cm^{-2}}$ is the total
absorption width  for the protons. 
It is possible to evaluate the width in a different way.
An estimate of the electron number 
density in the fireball is given in ref.\cite{Waxman:1997hj}: 
$n_{GRB}\simeq  (10^{10}\div 
10^{12})$ ${\rm  cm^{-3}}$. Using this value, and
taking the fireball mass in the range of star-like objects, 
$M = (1\div 10)\thinspace M_\odot$, we can get the radius of the object, $R_{GRB}
=5\cdot (10^{14}\div10^{15})$ cm, and then the width: 
$d_{GRB}=10\div 10^{4}~A$ ${\rm cm^{-2}}$. 
In agreement with the first argument, we see, then, that also in GRBs the matter
effect on neutrino conversion is negligible. 


\subsection{Dark matter halos}
 According to models\cite{Davis:1992ui}, 
part of the dark matter in halos of galaxies and clusters of galaxies
should consist of neutrinos\footnote{In what follows, we will not consider the
heavy particles present in the halos, because their number density is much
smaller than the one of neutrinos, although they provide the largest part of the mass of
the halo. Furthermore, the amplitude of the forward scattering of neutrinos on neutrinos and on the heavy
particles of dark matter are comparable, or the former is even larger.}.   
Therefore neutrinos of extragalactic origin crossing the halo on the way to the Earth undergo
refraction on the neutrino background.
It was suggested in ref.\cite{Horvat:1998ym} that, due to non uniform neutrino 
density distribution in the galactic halo, ultrahigh energy neutrinos can be resonantly 
converted into active and sterile species.

Following ref.\cite{Horvat:1998ym} we consider a galactic halo composed of 
non-relativistic neutrinos and antineutrinos of the two species 
$\nu_{\mu}$ and $\nu_{\tau}$.  The electron neutrino is assumed
to be lighter, and therefore less
clustered, than $\nu_{\mu}$ and $\nu_{\tau}$: $n_{\nu_e}\ll n_i$,  $i=\nu_{\mu},\nu_{\tau}$.
We assume CP-symmetry of the background: $n_i=n_{\bar i}$.
We take the density profile\cite{Horvat:1998ym}:
\begin{equation} 
n_\nu(r)=n^0_\nu {1\over 1+\left({r/a}\right)^2}~,
\label{eq:halo}
\end{equation}  
where  $n_\nu\equiv n_{\nu_\mu}+n_{\nu_\tau}$ and $a$ is the core radius of the halo. For  galactic
halos
this radius is estimated to be $a\simeq 10\div 100\thinspace{\rm kpc}$.  
An upper bound for $n^0_\nu$ is given by the
Tremaine-Gunn condition\cite{trem}: for identical fermions of mass $m$,
the maximum number density $n^{max}$ allowed by the
Pauli principle equals 
\begin{eqnarray} 
n^{max}={1\over 6\pi^2}\left( m v^{esc} \right)^3~,
\label{eq:tg}
\end{eqnarray}
where $v^{esc}$ is  the escape velocity of the particle:
\begin{eqnarray} 
v^{esc}=\left({2GM\over R}\right)^{1\over 2}\sim 540\left( {M\over { 10^{12}M_{\odot}}}\right)^{1\over 2}
\left( {R\over 30\thinspace {\rm Kpc}}\right)^{-{1\over 2}} \thinspace{\rm Km/s}.
\label{eq:vesc}
\end{eqnarray}
Here $M$ and $R$ are the total mass and radius of the galaxy.
In what follows we will take $R\simeq 3a$, corresponding to the radius 
at which the density reduces at one tenth of its core value, $n_\nu(3a)=n^0_\nu/10$.
From (\ref{eq:tg}) and (\ref{eq:vesc}) we get:
\begin{eqnarray} 
n^{max}
=1.7\cdot10^{6}\left( {m_\nu \over 5 \thinspace{\rm eV}}\right)^3
\left( {M\over {10^{12}M_{\odot}}}\right)^{3\over 2}
\left( {a\over 10\thinspace {\rm Kpc}}\right)^{-{3\over 2}}~{\rm cm^{-3}}. 
\label{eq:tgnum}
\end{eqnarray}  
The integration of the profile (\ref{eq:halo}) gives the matter width:
\begin{equation}
d={\pi \over 2} n^0_\nu a~.
\label{eq:dint}
\end{equation}
Inserting the expression (\ref{eq:tgnum}) for $n^{max}$ in (\ref{eq:dint}), we find the upper
bound for the width:
\begin{equation} 
d\lta 8\cdot 10^{28}\thinspace  \left( {m_\nu \over 5 \thinspace{\rm eV}}\right)^3
\left( {M\over {10^{12} M_{\odot}}}\right)^{3\over 2}\left( {a\over 10\thinspace 
{\rm Kpc}}\right)^{-{1\over 2}}{\rm cm^{-2}}~.
\label{eq:dgalhalo}
\end{equation}  
According to (\ref{eq:dgalhalo}) the largest values of 
 $d$ are achieved for objects
with big mass $M$ and small radius $a$; so that  compact halos 
represent the most favourable case. 

Let us now check the minimum width condition for 
$\nu_\mu-\nu_s$ and $\nu_\mu-\nu_e$ conversion in galactic halos.
We use the refraction width 
$d_0(s^{min}_Z)=0.055 d_0\simeq 1.35 \cdot 10^{31} {\rm cm^{-2}}$ 
given in eq. (\ref{eq:minimann}).  This value was obtained for $\nu_\mu-\nu_s$ conversion in
CP-symmetric neutrino background, and is the absolute minimum of $d_0(s_Z)$ (eq. (\ref{eq:dhalo})),
realized at the $Z$-boson resonance, $s_Z\sim 1$.
Notice that, under the assumption $n_{\nu_e}\ll n_i$, the result $d_0(s^{min}_Z)$ holds also for 
$\nu_\mu-\nu_e$ conversion: due to the absence of electron neutrinos in the background,
 $n_{\nu_e}\simeq 0$, the potential (\ref{eq:potnumusym}) for $\nu_e$ is negligible, and
therefore the electron neutrino behaves as a sterile species.

From eq. (\ref{eq:dgalhalo}) we see that, for typical values of $M$ and $a$ of a galaxy, like for
instance the Milky Way ($M\simeq 10^{12}M_{\odot}$ and $a\simeq 10 \thinspace{\rm kpc}$),
the minimum width condition is not satisfied: $d/d_0(s^{min}_Z)\lta 5\cdot 10^{-3}$.
For the galaxy M87  
($M\simeq 10^{13}M_{\odot}$ and $a\simeq 100 \thinspace{\rm kpc}$) we find 
$d/d_0(s^{min}_Z)\lta 5\cdot 10^{-2}$. 
Taking an hypotetical very massive and compact object, with
$M\simeq 10^{13}M_{\odot}$ and $a\simeq 10 \thinspace{\rm kpc}$ we get 
$d/d_0(s^{min}_Z)\lta 0.2$.
 Thus, a significant neutrino conversion effect in the galactic halo 
is excluded, in contrast with the result in ref.\cite{Horvat:1998ym}.
\\


Conversely,  significant matter-induced conversion can be realized in halos surrounding a
cluster of galaxies.  Taking the mass of a cluster as $M=(10^{13}\div 10^{15})M_{\odot}$
and the size $a\simeq 1 \thinspace {\rm Mpc}$, we obtain from eq. (\ref{eq:dgalhalo}) 
\begin{equation}
d\lta 10^{29}\div 3\cdot 10^{32}\thinspace{\rm cm^{-2}}\simeq (10^{-2}\div 20) d_0(s^{min}_Z)~.  
\label{eq:result}
\end{equation}
For the maximal value , $d/d_0(s^{min}_Z)\simeq 20$, from the condition
(\ref{eq:boundsin}) we get the sensitivity to the mixing:
$\sin^2 2\theta \gta 3\cdot 10^{-3}$.  With this value we find that  
the adiabaticity condition (\ref{eq:gamma}) is fulfilled for $\Delta m^2\gta 4\cdot 10^{-6}~{\rm
eV^2}$.

The maximal sensitivity is achieved in the energy range of the $Z$-resonance, where also inelastic
scattering and absorption are important.  Indeed, in section 4.4, taking $\sin 2\theta=0.3$, we have 
found that $d_{abs}\sim d_{min}$ at $s\simeq 6\cdot 10^{3} ~{\rm GeV^2}$, where  
$d_{min}\simeq 3 \cdot 10^{32} {\rm cm^{-2}}$. This value coincides with 
the upper edge of the interval (\ref{eq:result}).
For smaller $\sin 2\theta$ $d_{min}$ is larger, and the absorption effect on oscillations becomes even
more important.  Notice that $d_0(s_Z)$ takes its minimum value at the $Z$-resonance: for neutrino 
energies outside the resonance $d_0(s_Z)$ is larger, so that $d/d_0(s_Z)<1$ and the minimum width
condition is not satisfied.

Thus, we have found that in the halos of clusters of galaxies  a significant matter-induced conversion
can  be achieved  in the narrow interval of energies of the $Z$-boson resonance, where, however, the
absorption and the effects of inelastic interactions are important. 


\subsection{Early Universe}
In this section we consider neutrinos produced by cosmologically distant sources and
propagating in the universe.  The refraction occurs due to the 
interaction of the neutrinos with the particle
background of the universe made of neutrinos electrons and nucleons.

The number densities
of baryons, $n_b$, and electrons, $n_e$, are given by 
$n_b=n_e=\eta_b n_\gamma$, where 
$\eta_b=10^{-10}\div10^{-9}$ is the baryon asymmetry of the universe and 
$n_\gamma$ is the concentration of photons.  At present time  $n_\gamma=n^0_\gamma\simeq 400~{\rm cm^{-3}}$.  
We will describe the neutrino background by the total number density $n_\nu+n_{\bar{\nu}}$ and the
CP-asymmetry $\eta_{\nu}\equiv (n_\nu-n_{\bar{\nu}})/n_\gamma$.  The value of $\eta_{\nu}$ is unknown.
A natural assumption would be $\eta_{\nu}\simeq \eta_b$: in this $-$almost
CP-symmetric$-$ case the total concentration of neutrinos equals 
$n_\nu+n_{\bar{\nu}}=4 n_\gamma /11$\cite{kolb}.  However strong asymmetry, $\eta_{\nu}\sim 1$, is not
excluded. The upper bound $\eta_{\nu}\sim 10$ for muon and tau
neutrinos follows from the Big Bang Nucleosinthesis\cite{Kang:1992xa,Lesgourgues:1999ej}.  

For $\eta_{\nu}\sim 1$ the contribution of the neutrino background to refraction
dominates and the interaction of
neutrinos with electrons and nucleons can be neglected. 
It can be checked that the contributions of
neutrino-electron and neutrino-nucleon scattering to the refraction width are
smaller than $d_0$ at any time after the neutrino decoupling epoch, $t_{dec}\simeq 1$ s.
\\
\\



In the framework of the standard Big-Bang cosmology, the number density of
neutrinos in the universe decreases with the increasing time as\cite{kolb}:  
\begin{equation}
n(t)=\cases{n_0\displaystyle{\left({t_0\over t} \right)^{2}}
 & $t\geq t_{eq}$ \cr 
n_{eq}\displaystyle{\left({t_{eq}\over t} \right)^{3\over 2}} & $t< t_{eq}$ ~.\cr}
\label{eq:neev} 
\end{equation} 
Here $t_{0}\simeq 10^{18}$ s is the age of the universe, and 
$t_{eq}\simeq 10^{12}$ s is the time at which the energy densities of radiation and of
matter in the universe were approximatively equal.
We denote by $n_0$ and $n_{eq}$ the neutrino concentrations at $t=t_0$ and $t=t_{eq}$
respectively.

The matter width  
$d(t)$ crossed by the neutrinos from the time $t$ of their
production to the present one is given by the integration of the concentration 
(\ref{eq:neev}):
\begin{equation}
d(t)=\int^{t_{0}}_t n(\tau) d\tau
=\cases{d_{U}\displaystyle{\left[{t_0\over t} -1\right]}
&~~ $t\geq t_{eq}$\cr 
d_{eq}\displaystyle{\left({t_{eq}\over t} \right)^{1\over 2}}  &~~ $t<t_{eq}$~,\cr}
\label{eq:dtev} 
\end{equation}
where $d_{U}\equiv t_0 n_0$ is the present width of the universe and 
$d_{eq}=d_{U}\left[{t_0/ t_{eq}} -1\right]$ is the width at $t=t_{eq}$.

In what follows we will focus on the case of matter domination epoch, 
$t\geq t_{eq}$, for which the width (\ref{eq:dtev}) can be expressed in terms of the redshift,
$z\equiv\left({t_0/ t}\right)^{2/3}-1$, as:
\begin{eqnarray}
d(z)=d_{U}\left[(z+1)^{3\over 2}-1\right]=
d_{i}\left[1-(z+1)^{-{3\over 2}}\right]~, 
\label{eq:dz} 
\end{eqnarray}
where $d_i=t n=d_{U}(z+1)^{3\over 2}$ is the width of the universe at the time $t$ of production
of the neutrino beam; $n$ is the concentration of the neutrino background at $t$. 
According to eq. (\ref{eq:dz}),  for large enough $z$ the width at the production time, $d_i$, 
gives the dominant contribution.  

Another important feature of the propagation of neutrinos from cosmological sources is the redshift
of energy. The refraction width $d_0$ depends on the center of mass energy squared  $s$
of the incoming and the background neutrinos. As a consequence of the redshift, 
$s_Z=s_Z(z)$, the width $d_0$ changes with time (with $z$) during the neutrino propagation: 
$d_0=d_0(s_Z(z))$.  Thus, the width of matter $d$ should be compared with some effective (properly
averaged) refraction width $d_0$ which in fact depends on the channel of transition. 
For non-relativistic background neutrinos of mass $m_\nu$  
we have that $s$ increases with $z$ as:
$s\simeq 2m_\nu E \propto (1+z)$, where $E$ is the energy of the neutrino beam. 
For relativistic background with energy $E_b$ one gets $s\simeq 2 E_b E \propto (1+z)^{2}$.
\\

Let us consider neutrino propagation in a strongly CP-asymmetric background, $n_i\gg n_{\bar i}$, with 
$\eta_\nu\sim 1$.  For simplicity we assume also flavour symmetry:
$n_{\nu_e}=n_{\nu_\mu}=n_{\nu_\tau}$.

For the $\nu_\alpha - \nu_s$ channel  the refraction width 
(\ref{eq:dun}) increases smoothly from its low energy value, $d_0(s_Z\ll 1)=3d_0/4$, to the high
energy one, $d_0(s_Z\gg 1)=d_0$. For neutrinos produced with energy  $E\lta 10^{21}$ eV and mass of
the background neutrino $m_{\nu_\alpha}\simeq 2$ eV we get $s_Z\lta 0.5$, which undergoes redshift
during the neutrino propagation.   Therefore we can use the low energy value of $d_0$ as the effective
refraction width.  
From (\ref{eq:dz}) we get  the ratio $r\equiv {d/ d_0(s_Z)}$ 
in terms of the redshift $z$ and the  asymmetry $\eta_\nu$:
\begin{eqnarray}
r(z)\equiv {4d(z)\over 3 d_0}\simeq 2.2\cdot 10^{-2}\eta_\nu(z+1)^{3\over 2}~.
\label{eq:dasym} 
\end{eqnarray}
Taking the maximal allowed asymmetry, $\eta_{\nu}\sim 10$, we find that $r=3$ is reached at $z\simeq 5$,
which corresponds to rather recent epoch. Possible sources of high energy neutrinos, the quasars, have
been observed at such values of the redshift.
With smaller asymmetries, $\eta_\nu \lta 1$, the minimum width condition requires much earlier  
epochs of neutrino production, $z\gta 27$.

Notice that in general the minimum width condition, $d\geq d_0/\sin 2\theta$, is not sufficient to
ensure a significant transition effect: as discussed in section 2, the width $d_{1/2}$ 
required to have conversion probability $P\geq 1/2$ depends on the specific effect involved. For 
the adiabatic conversion in varying density (section 2.3) we have found  the result 
$d_{1/2}\simeq 1.5 d_{min}$ (see eq. (\ref{eq:mind12lz})).  
Therefore, the condition $P\geq 1/2$ requires
larger values of $r(z)$, $r\gta 4.5$.  From eq. (\ref{eq:dasym}), with  
$\eta_{\nu}\sim 10$, we get $r\sim 4.5$ for $z\simeq 7.5$. 
\\

For $\bar{\nu}_\alpha- \bar{\nu}_s$ channel the refraction width $d_0$, eq. (\ref{eq:dunanti}), 
has a resonance character with absolute minimum $d_0(s^{min}_Z)\simeq d_0/7$ at $s^{min}_Z\sim 1$.
Outside the resonance it takes the values  $d_0(s_Z\ll 1)=3d_0/4$ and $d_0(s_Z\gg 1)=d_0$. 
For neutrino  energy  $E\lta 10^{21}$ eV at production and mass of
the background neutrino $m_{\nu_\alpha}\lta 1$ eV we get $s_Z< 1$, so that we can use the low energy
value of the refraction width and the result for the ratio $r$ coincides with that in eq.   
(\ref{eq:dasym}).  For neutrinos produced at time $t=t_i$ with extremely high energies, 
$E\sim  10^{21} \div 10^{22}$ eV and mass of the background neutrino  
$m_{\nu_\alpha}\simeq 1\div 3 $ eV we get $s_Z\geq 1$ at the
production time, so that, due to redshift, $s_Z(z)$ will cross the resonance interval, in which $d_0$
has minimum.  This, however, does not lead to larger values of the ratio $r(z)$, since the time interval
$\Delta t$ during which $s_Z$ remains in the resonance range is short: 
$\Delta t/t_i\simeq 1.5 \gamma_Z\simeq 0.04$.  The matter width collected in the
interval $\Delta t$ is $d_{res}\simeq d \Delta t/t_i\simeq 1.5 \gamma_Z d$, 
and the ratio $r(z)$ for the resonance epoch equals $r(z)=d_{res}/d_0(s^{min}_Z)\simeq 
d/4 d_0$. That is even smaller than $r$ outside the resonance, eq. (\ref{eq:dasym}); therefore
the result (\ref{eq:dasym}) holds also in this case.

Thus,  in the extreme condition of very large $\nu-\bar{\nu}$ 
asymmetry and production epoch at $z\gta 5$ the matter width crossed by neutrinos satisfies 
the minimum width condition.  Let us comment on the character of the matter-induced neutrino
conversion in this case.  After its production, the neutrino beam experiences a monotonically
decreasing density.  Moreover, the energy of neutrinos decreases due to redshift, which also
influences the mixing.  
Taking into account the decrease with time of both the neutrino energy and concentration, the
adiabaticity condition (\ref{eq:gamma})-(\ref{eq:adiab}) can be generalized as:
\begin{eqnarray}
&&\gamma(t)=\gamma_0 {t \over t_0}=\gamma_0 (z+1)^{-{3\over 2}}\ll 1 
\label{eq:gammat} \\
&&\gamma_0\simeq {8 \over 3 V_0 t_0 \tan^2 2\theta}~,
\label{eq:adiab0} 
\end{eqnarray}
where $V_0$ is the present value of the neutrino-medium potential.   
For the present epoch we get $\gamma_0\simeq 10^2/\tan^2 2\theta$, so that the adiabaticity is
strongly broken.  For $\tan^2 2\theta \lta 0.1$ the adiabaticity can be realized at
$t/t_0<10^{-3}$, or $z>10^2$.   Thus, for $z<10^2$ the conversion takes a character of
oscillations in the production epoch.  
A detailed study of the dynamics of
neutrino conversion in the universe will be given elsewhere\cite{next}.
\\

For the flavour channel  $\nu_\alpha - \nu_\beta$ the refraction effect appears at high energies,
$s_Z\sim 1$, as a consequence of different energies of the background neutrinos
$\nu_\alpha$ and $\nu_\beta$. The absolute minimum  of the refraction width (\ref{eq:dunab}), 
$d^{min}_0(s^{i}_Z )=3 d_0$, gives a value of $r(z)$ which is 4 times smaller than that in eq. 
(\ref{eq:dasym}).  Correspondingly, even for the extreme conditions of $\eta_\nu\simeq 10$ and 
$z=5$ we get $r\simeq 0.8$. That is, significant matter effect is excluded for neutrinos from the
oldest observed sources. The value $r\gta 3$ can be reached for $z\gta 14$.
Notice that, according to eq. (\ref{eq:dunab}), the refraction disappears for low energies, $s^i_Z\ll
1$, and the redshift spoils the conditions of absolute minimum of $d_0$, even if it is realized 
in certain epoch.  Therefore, larger $z$ are required to have significant conversion effect.

Similar conclusions can be obtained for $\bar{\nu}_\alpha - \bar{\nu}_\beta$ channel, where the
refraction width  (\ref{eq:dunabCP}) has the  local minimum $d_0(s^{i}_Z)\simeq 6 \gamma_Z d_0$
due to $Z$-resonance.  This minimum is realized however during the resonance epoch 
$\Delta t$. Taking the  corresponding  matter width  $d_{res}\simeq  1.5 \gamma_Z d$, 
we get $r(z)=d(z)/4 d_0$, which is even smaller than in the $\nu_\alpha - \nu_\beta$ case. 
\\

Let us consider CP-symmetric neutrino background, $n_i=n_{\bar i}$, 
with the assumption of flavour symmetry: $n_{\nu_{e}}=n_{\nu_{\mu}}=n_{\nu_{\tau}}$.
As discussed in section 4.4, unsuppressed refraction effect appears for active-sterile neutrino
channels at high energies, $s_Z\sim 1$, where the propagator corrections become important.

For $\nu_\alpha - \nu_s$ (and similarly for $\bar{\nu}_\alpha- \bar{\nu}_s$, due to CP-symmetry) the
refraction width (\ref{eq:dhalo}) has the absolute minimum  
$d_0(s^{min}_Z)\simeq 2\gamma_Z d_0$, realized at $s^{min}_Z=1\pm \gamma_Z$, eq. (\ref{eq:minimann}).
Assuming that the neutrinos are produced just before the resonance epoch, we find that the width 
collected during the interval $\Delta t$ equals  $d_{res}\simeq d \Delta t/t_i\simeq 1.5 \gamma_Z d$, 
where $d$ is given by eq. (\ref{eq:dz}), with $d_U=2 n^0_\gamma t_0/11 \simeq 7 \cdot 10^{29}~{\rm cm^{-2}}$.
For the ratio $r(z)$ we find:
\begin{equation}
r(z)={3 d(z)\over 4 d_0}=2\cdot 10^{-3}(z+1)^{3\over 2}~,
\label{eq:rzcpsym}
\end{equation}
which is significantly  smaller than $r(z)$ for $\nu_\alpha - \nu_s$ channel in
CP-asymmetric background, eq. (\ref{eq:dasym}).   With the maximal
redshift $z\simeq 5$ eq. (\ref{eq:rzcpsym}) gives $r\simeq 0.03$, which excludes matter effect for
neutrinos from the most distant observable sources.
The result (\ref{eq:rzcpsym}) holds also for the $\nu_\alpha - \nu_\beta$ ($\bar{\nu}_\alpha - \bar{\nu}_\beta$)
channel, since the refraction width (\ref{eq:dabcpsym}) has analogous behaviour to the one for the
$\nu_\alpha - \nu_s$ case, eq. (\ref{eq:dhalo}), with the same local minimum  $d_0(s^{i}_Z)\simeq 2 \gamma_Z d_0$
at resonance. 
\\

In conclusion, we have found that the matter effect for neutrinos crossing the universe is mainly due
to the neutrino background.  
For neutrinos from observable sources ($z\simeq 5$),
significant conversion effect can be achieved  in the $\nu_\alpha - \nu_s$ and 
$\bar{\nu}_\alpha-\bar{\nu}_s$ channels, 
if the background has 
strong CP-asymmetry, close to the maximum value,
$\eta_\nu\simeq 10$. 
The matter effect
for the other conversion channels and for the CP-symmetric case is suppressed as a consequence
the redshift.


\section{Conclusions}

\noindent
$1)$. Matter effects can lead to strong flavour transition even for small vacuum mixing angle:
$\theta\ll 1$.  This however requires a sufficiently large amount (width) of matter
crossed by neutrinos: the minimum width
condition, $d\geq d_{min}$, should be satisfied, where 
$d_{min}= \pi /(2\sqrt{2} G_{F}\tan 2 \theta)=d_0/\tan 2\theta$, 
for low neutrino energies, $s\ll M^2_Z$, and conversion probability $P\geq 1/2$.
The absolute minimum $d_{min}$ is realized for uniform medium with resonance density 
$n^{res}_e$.
\\

\noindent
$2)$. We have shown that for all the other realistic situations the required width, $d_{1/2}$,
is larger than $d_{min}$.  In particular, we have found that 
$d_{1/2}/d_{min}=1+(1-\pi/8)\gamma^2$ for oscillations in medium  with slowly varying density 
($\gamma\ll 1$); $d_{1/2}/d_{min}\geq 1.5$ for conversion in medium with varying density;
$d_{1/2}/d_{min}= \pi$ for castle wall profile. 
\\

\noindent
$3)$. We discussed the minimum width condition for  high energy neutrinos.
For $s\gta M^2_W$ the minimum width $d_{min}$ becomes function of $s$, due to the
propagator effect: $d_{min}=d_{min}(s)$. The function $d_{min}(s)$ depends on the channel of
interactions: in the case of $W$ (or $Z$) exchange in the  $s$-channel $d_{min}(s)$ decreases 
in the resonance region by a factor $\sim 20$ with respect to the low energy value: 
$d_{min}(0)/d_{min}(M^2_W)\sim 20$. In this region, however, the inelastic interactions become
important, damping the flavour conversion. 
\\

\noindent
$4)$. As a case of special interest we have studied the refraction of high energy neutrinos in
neutrino background, which can be important for propagation of cosmic neutrinos in galaxies and
intergalactic space.  
Again we find that the $\nu_{\alpha}-\bar\nu_{\alpha}$ annihilation channel gives enhancement 
of refraction at $s\simeq M^2_Z$, so that $d_{min}$ can be $\sim 1/2\gamma_Z\sim 20$ times
smaller than that  at low energies.  
In the case of flavour channels the refraction can appear as the result of the difference of masses of
the background neutrinos, even if the concentrations of the various flavours are equal. 
\\

\noindent
$5)$. The minimum width condition allows one to conclude on the relevance of the matter effect 
without knowledge of the density profile, once the width $d$ is known.
In some astrophysical situations the total width on the way of neutrinos can be  
estimated rather precisely (e.g. by spectroscopical methods) although the density distribution
is unknown.  Significant matter effect in excluded if $d<d_0$, or $d<d_{min}$ if the mixing
angle is known. 
\\

\noindent
$6)$. From practical point of view, a study of the matter effects should start with the check
of the minimum width condition, $d\geq d_{min}$.  This condition is necessary but not sufficient
for strong conversion effect.  If it is fulfilled   
the ratio $d/d_0$ allows one to estimate the minimal
mixing angle for which significant transition is possible: $\sin 2\theta>d_0/d $. This condition
gives an absolute lower bound on $\theta$, which can be achieved for the case of uniform medium 
with resonance density. 
In other words, given the width $d$ of the medium, the highest sensitivity
to the mixing angle is achieved if the matter is distributed uniformly and the density
coincides
with the resonance value for a given neutrino energy. 
For media with non-uniform matter distribution the sensitivity to $\theta$ is lower.
The stronger the deviation from the constant density, the lower the sensitivity. 
\\

\noindent
$7)$. We applied the minimum width condition to neutrinos in AGN and GRBs environment.
For AGN the width $d$ can be estimated by the experimental data on the X-ray spectrum, without
assumptions on the the density profile. We got $d/d_0\lta 10^{-10}$ for radially moving neutrinos, 
strongly excluding matter effects.  
In the case of GRBs the width $d$ can be evaluated under the assumption that the object is
transparent to protons.
We found $d/d_0\lta 10^{-5}$. Therefore, no significant
conversion is expected either.
\\

\noindent
$8)$. For  neutrinos crossing the
halos of galaxies and clusters of galaxies 
the matter effect is given
by the interaction of neutrinos with the neutrino component of the halo.
We have found that for galactic halos the minimum width condition is not
satisfied: the result $d(halo)/d_0 \leq 0.1$
excludes any significant conversion effect.
For halos of clusters of galaxies we got $d(halo)/d_0 \gta 10$, and the minimum width condition
can be satisfied for large enough mixing: $\sin2\theta \gta 0.1$. 
\\

\noindent
$9)$.  
We have considered the refraction of neutrinos from cosmologically distant sources, interacting with
the neutrino background of the universe. 
Significant active-sterile conversion can be expected in case of large $\nu-\bar{\nu}$
asymmetry.  We have found that for 
$\eta_{\nu}=O(1)$ the condition $d(universe)/d_0\gta 1$ can be achieved for neutrinos from
sources, galaxies or quasars, with redshift $z\gta 5$.
The effect on detected neutrino fluxes from these sources could be a distortion of the energy
spectrum.

\vspace{1cm}
\noindent
{\bf Acknowledgments} 
 
\noindent
The authors wish to thank O.L.G.~Peres for useful comments.


\bibliography{width}
\end{document}